\documentclass{pazhb_eng} 
     
\usepackage{graphicx} 
\usepackage{amsmath} 
\usepackage{epsfig,graphics} 
\usepackage{rotating} 
\usepackage{ulem}

\usepackage{epstopdf}
\usepackage{pdfpages}

\usepackage{longtable}
\usepackage{dcolumn}
\usepackage{pdflscape}

\usepackage{hyperref}

\usepackage{comment}
\usepackage{ulem}
\usepackage{color}
\usepackage{tcolorbox}
\usepackage{subcaption}
\usepackage{ amssymb }

\usepackage{ulem}
\usepackage{floatpag}
     
\newcommand {\be}{\begin{equation}} 
\newcommand {\ee}{\end{equation}}

\newcommand{\Lagr}{\mathcal{L}}

\def\red#1{\textcolor{red}{#1}}
\def\БР#1{\red{\textbf{#1}}}

\newcommand{\zspec}{z_{\mbox{\scriptsize spec}}}
\newcommand{\zphot}{z_{\mbox{\scriptsize phot}}}

\def\2MASS{{2MASS}}
\def\WISE{{WISE}}
\def\3XMMDR4{{3XMM-DR4}}
\def\SDSS{{SDSS}}

\def\XMM{{XMM}-Newton}
\def\BTA{BTA}
\def\AZT33IK{АZT-33IK}

\def\K16{{K16}}
\def\V14{{V14}}
\def\V14U{{V14U}}
\def\XMMXXL{{XMM-XXL}}

\def\kev2l{L_{\rm 2\,keV}}
\def\l2500A{L_{2500\mathring{A}}}
\def\LXLO{$\kev2l$--$\l2500A$}
\def\LX210{L_{{\rm X},2-10}}

\def\FX0.52{F_{{\rm X},0.5-2}}

\def\detML{det_{\rm ML}}
\def\Ndet{N_{\rm det}}
\def\Vmax{V_{\rm max}}

\def\zmin{z_{\rm min}}

\def\elum{e_{\rm lum}}
\def\eden{e_{\rm den}}
\def\plum{p_{\rm lum}}
\def\pden{p_{\rm den}}
\def\NH{N_{\rm H}}
\def\zmag{\mbox{z}^\prime}
\def\rmag{\mbox{r}^\prime}

\def\gamaod{2.78}
\def\gamaoderr{0.20}

\def\CmdA2{/home/george/Documents/QSO3LumfuncArticle/Plot/FR_Plot_textableinbins/CmdA2/}
\newcommand*{\PathtoIMAGE}{IMAGES}

\newcommand{\Azt}{АЗТ-33ИК}

\newcommand{\Nmylumphot}{11}
\newcommand{\Nmysamplelum}{101}
\newcommand{\Nmyvito}{104}
\newcommand{\Nmylumtot}{205}
\newcommand{\Nmyspeccnt}{8}
\newcommand{\Nspeccnt}{90}

\hypersetup{draft}

\begin{document}

\journalinfo{2018}{44}{8-9}{500}[521]
\UDK{524.7}


\title{X-ray Luminosity Function of Quasars at 3<\mbox{\small
    z}<5 from XMM-Newton Serendipitous Survey Data}


\author{G.~A.~Khorunzhev\email{horge@iki.rssi.ru}
  S.~Yu.~Sazonov, R.~A.~Burenin
  \vskip -2mm ~\\
  {\rm\em\small  Space Research Institute, Russian Academy of Sciences, Profsoyuznaya ul. 84/32, Moscow, 117997 Russia}
}

\shortauthor{Khorunzhev~\etal}  
\shorttitle{X-ray Luminosity Function of Quasars at $z>3$} 

\submitted{10.11.2017 г.}

\begin{abstract}  
   The X-ray luminosity function of distant ($3<z<5.1$)
   unabsorbed quasars has been measured.  
   A~sample of distant high-luminosity quasars ($10^{45} \leq \LX210 < 7.5 \times 10^{45}$~erg/s
   in the 2--10~keV energy band) from the catalog given in Khorunzhev~et~al.~(2016) 
   compiled from the data of the {3XMM-DR4} catalog 
   of the {XMM}-Newton serendipitous survey and the Sloan Digital Sky Survey (\SDSS) has been used.  
   This sample consists of \Nmysamplelum \ sources. 
   Most of them (\Nspeccnt) have spectroscopic redshifts $\zspec\geqslant 3$.
   The remaining ones are quasar candidates with photometric redshift estimates $\zphot\geqslant 3$. 
   The spectroscopic redshifts of eight sources have been measured with \AZT33IK and \BTA \ telescopes.
   Owing to the record sky coverage area ($\simeq 250$~sq.~deg at 
   X-ray fluxes $\sim 10^{-14}$~erg/s/cm$^{2}$  in the 0.5-2~keV),
   from which the sample was drawn, we have managed to obtain 
   reliable estimates of the space density of distant
   X-ray quasars with luminosities 
   $\LX210 > 2 \times 10^{45}$~erg/s for the first time.  
   Their comoving space density remains constant as the redshift increases from $z=3$ to $z=5$ 
   to within a factor of 2. The power-law slope
   of the X-ray luminosity function of high-redshift quasars 
   in its bright end (above the break luminosity) has been reliably constrained for the first time.
   The range of possible slopes for the quasar luminosity dependent density evolution
   model is $\gamma_2=2.78^{+0.00}_{-0.04}\pm0.20$, where initially the lower and upper boundaries of $\gamma_2$ 
   with the remaining uncertainty in the detection completeness of X-ray sources in \SDSS, and subsequently the
   statistical error of the slope are specified.
 
   
\end{abstract}

\section{Introduction}

A reliable measurement of the X-ray luminosity
function of high-luminosity active galactic nuclei
(AGNs, hereafter quasars) and its evolution at
 $z\gtrsim 3$ is one of the most important components
of the research on the growth history of supermassive
black holes and the evolution of massive
galaxies in the Universe. The samples of \XMM\ and
{Chandra} extragalacitc X-ray surveys (representative fluxes $\FX0.52\lesssim~10^{-15}$~erg/s/cm$^2$ and areas about one~sq.~deg) turn out to be insufficiently large
for the evolution of distant quasars to be studied \citep{civano12, vito14}.
The addition of sources
from shallower extragalactic surveys ($\FX0.52\sim 10^{-14}$--$10^{-13}$~erg/s/cm$^2$) covering much larger areas  (tens of square degrees \citealt{ueda14,aird15,georgakakis15}) improves the situation.

\cite{vito14} constructed and extensively
studied the luminosity function of quasars at \mbox{$z>3$} with luminosities $\LX210<10^{45}$~erg/s in the 2--10~keV band based on the combined data of several
deep X-ray surveys with a total area \mbox{$\simeq 3.3$~sq.~deg}. Using data from the {XMM-XXL} survey with an area of 18~sq.~deg
(typical fluxes of sources $\FX0.52\simeq5\times10^{-15}$~erg/s/cm$^2$, \citealt{menzel16}),   \cite{georgakakis15}, obtained statistically significant estimates of the quasar luminosity funcion at $z>3$ for even higher luminosities ($\LX210\gtrsim 10^{45}$~erg/s).

\cite{ueda14} studied the evolution of the
X-ray luminosity function of AGNs based on the
collection of data from a large set of X-ray surveys,
including the {ROSAT} all-sky survey. The {ROSAT}
sample of sources includes several quasars with a
very high luminosity ($\LX210>10^{46}$~erg/s) at $z>3$, which allowed the space density of such very luminous and 
distant quasars to be constrained.
This estimate turned out to be in agreement with the predictions of
the empirical luminosity function model obtained from samples of sources with a much lower
luminosity ($L_{X,2-10}<10^{45}$~erg/s). 

\cite{kalfonzou14} compiled a catalog of
quasars at $z>3$ on an area of $\simeq 33$~sq.~deg based on
the archival data of individual nonoverlapping Chandra
pointings over the entire time of its operation.
Using this catalog, they were able to estimate the
space density of distant quasars with luminosities $L_{X,2-10}>5\times 10^{44}$~erg/s and to exclude some of
the empirical luminosity function models. However, the size of this sample
is still insufficient for a detailed study
of the population of most luminous \mbox{($\LX210>10^{45}$~erg/s)} and distant ($z>3.5$) quasars.  

The data from the XMM-Newton X-ray telescope
accumulated over 15 years constitute a serendipitous
sky survey \citep{watson09} with a total coverage of $\sim$800~sq.~deg and a sensitivity $\FX0.52\sim 5\times 10^{-15}$~erg\,s$^{-1}$\,cm$^{-2}$ \citep[the {3XMM-DR4} fourth data release of serendipitous
   source catalog\footnote{\url{http://heasarc.gsfc.nasa.gov/W3Browse/xmm-newton/xmmssc.html}},][]{watson09}. 

Based on these data,
one can produce an X-ray sample of quasars at $z>3$ that exceeds the existing samples by several times \citep{kalfonzou14,georgakakis15} and obtain more rigorous constraints on 
 the luminosity function model parameters. This is the goal of our paper.

We made an attempt to find new candidates for distant quasars
among the X-ray sources of the {3XMM-DR4} catalog as described in \citep{khorunzhev16,khorunzhev17}.
Our goal was to obtain a sample of X-ray quasars
at $z>3$ as complete as possible in {XMM}-Newton serendipitous survey fields at Galactic latitudes $|b|>20^\circ$
using photometric data from the Sloan Digital Sky Survey \citep[\SDSS,][]{alam15} as well as the infrared
{\2MASS} \citep{cutri03} and {\WISE} \citep{wright10}. The total area of the
overlap between these surveys is 300~sq.~deg.

The photometric
redshift estimates ($\zphot$) had been done by \cite{khorunzhev16AA} and a catalog
(\K16) of 903 candidates for distant quasars (presumably
of type 1) selected by photometric redshift had been compiled.
The catalog includes both previously known quasars
(with measured spectroscopic redshifts $\zspec>3$) and
new quasar candidates (with photometric redshift estimates
$\zphot>2.75$).

The additional table of the
\K16 \ catalog presents 63 known X-ray quasars with
$\zspec>3$ that did not pass the photometric selection
of quasar candidates. The first results of our
spectroscopic identification of new quasar candidates
from the \K16 catalog, based on which we made a
quantitative estimate of the purity of this catalog,
are presented in \cite{khorunzhev17}, \cite{khorunzhev17b}.
The additional selection was shown to provide an increase
in the number of new sources at $z>3$ relative
to the existing spectroscopic sample of quasars: by
$\sim 20$\% for optically bright ($\zmag<20$) and X-ray ($\LX210\gtrsim 10^{45}$~erg/s)
luminous sources and by $\sim 50$\% for fainter sources.

In this paper we use data from the K16 catalog
to measure the space density of luminous ($\LX210>10^{45}$~erg/s) quasars at $z>3$ 
and to obtain rigorous
constraints on the slope of the luminosity function
$\gamma_2$ in its bright end. In our calculations we used
the following cosmological constants, the same as
those in Vito~et~al.~(2014), whose results are actively
used below: $H_0=70$~km/s/Mpc, $\Omega_{\rm m}=0.27$, $\Omega_\lambda=0.73$.

\section{THE SAMPLE}\label{sec:vyborka}

To construct the X-ray luminosity function, we
used a sample of \Nmylumtot\ sources composed of the
parts of two catalogs: \Nmysamplelum\ sources with luminosities
$\LX210\geq 10^{45}$~erg/s from the catalog by \cite{khorunzhev16} and \Nmyvito \ unabsorbed sources with
$\LX210<1.1\times10^{45}$ erg/s from the catalog by \cite{vito14}.

\subsection{The Subsample of Luminous Quasars from the \K16 Catalog}

To investigate high-luminosity ($\geq 10^{45}$~erg/s)
quasars, we used the \K16\ catalog of quasars and candidates
for distant quasars \citep{khorunzhev16}.
We considered both objects from the main catalog
and sources from the additional table of known
quasars with $\zspec >3$ that did not pass the photometric selection.
The sources that were \XMM  \
pointing targets and the blazar
3XMM~J142437.8+225601 were excluded.

\begin{figure}[!ht]
\centering
\includegraphics[width=\linewidth]{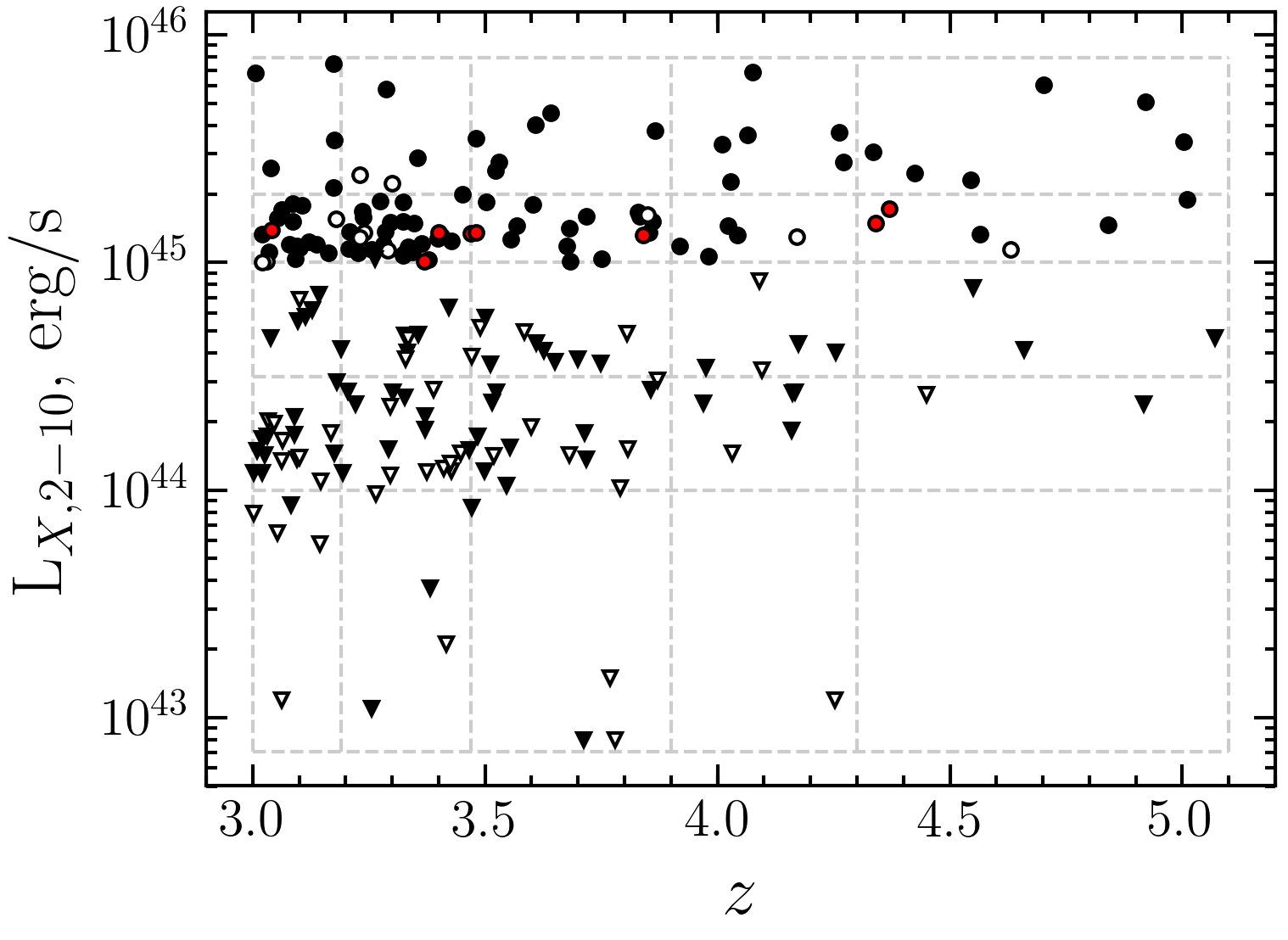}

\centering
\includegraphics[width=\linewidth]{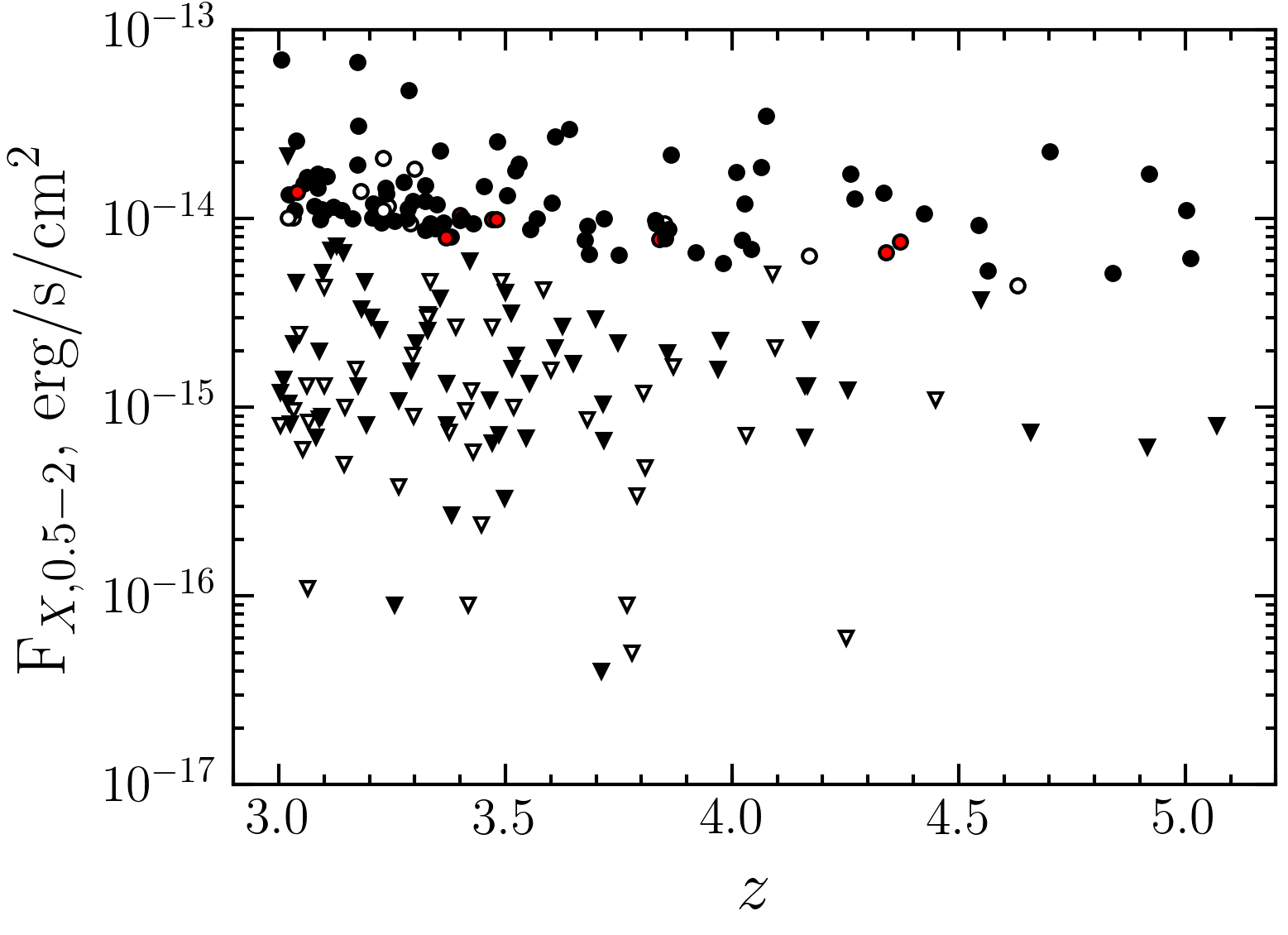}
\caption{Top: X-ray luminosities (2--10~keV in the objects's rest frame) and redshifts of the quasars from the \K16\ (circles) and \V14U\ (triangles) subsamples. 
The vertical and horizontal dashed lines indicate the boundaries of the $\Delta\log L$--$\Delta z$ bins
for constructing the binned (nonparametric) luminosity function by the $1/\Vmax$ method. Bottom: X-ray fluxes (0.5--2~кэВ) and 
redshifts of the sources from the same subsamples. 
The filled symbols indicate the objects with spectroscopic redshifts ($\zspec$).
The objects
from the \K16 catalog whose spectroscopic redshifts were measured already after the publication of the catalog \citep{khorunzhev17,khorunzhev17b,khorunzhev18b} are highlighted by the red color. 
The open symbols indicate the objects for which there are only photometric redshift estimates ($\zphot$).} 
\label{fig:catflux}
\end{figure}

As a result, we selected \Nmysamplelum \ sources with~2–10~keV X-ray
luminosities $\LX210
\geq10^{45}$~erg/s in the source’s
rest frame. The luminosity was calculated via the
measured 0.5-2~keV flux under the assumption of
a power-law spectrum with a slope $\Gamma$=1.8 without
absorption (just as in \citealt{vito14} for unabsorbed sources). 
In the case where an object had no spectroscopic
redshift, the luminosity was calculated from
$\zphot$, the photometric redshift estimate. As a result
of the selection by luminosity, all sources of the K16
subsample turned out to have an X-ray flux above 4$\times 10^{-15}$~erg/s/cm$^2$. 
The distribution of sources in
X-ray flux, luminosity, and redshift is shown in~Fig.~\ref{fig:catflux}.
The list of sources is given in Table~\ref{tab:catlum}.

For 82 of the \Nmysamplelum \ sources $\zspec \geq 3$ was known at
the time of \K16 publication. The sample also includes
\Nmyspeccnt \ spectroscopically confirmed candidates with
$\zspec \geq 3$ whose spectra were taken with the 1.6-m \Azt\ telescope at the Sayan Solar Observatory
of the Institute of Solar–Terrestrial Physics,
the Siberian Branch of the Russian Academy of
Sciences, and the \mbox{6-m} \BTA \ telescope at the Special
Astrophysical Observatory of the Russian Academy
of Sciences during our program of searching for
distant quasars \citep{khorunzhev17,khorunzhev17b,khorunzhev18b}.
The remaining \Nmylumphot \ objects are quasar candidates
with photometric redshift estimates $\zphot \geq 3$
unambiguously identified in the optical band
(without the "D"\ flag in the  \K16 catalog).
The source
3XMM~J114816.0+525900 ($\zspec =3.173$) has the
highest luminosity $\LX210 =7.4\times 10^{45}$~erg/s.
3XMM~J022112.5-034251 is the most distant source
$\zspec=5.011$, \mbox{$\LX210 =1.9\times 10^{45}$~erg/s.}
Our sample contains several times more X-ray luminous
quasars than the previously used data from smaller area
X-ray surveys \citep{kalfonzou14,vito14,aird15,georgakakis15}.

\begin{figure}[!ht]
\centering
\includegraphics[width=\linewidth]{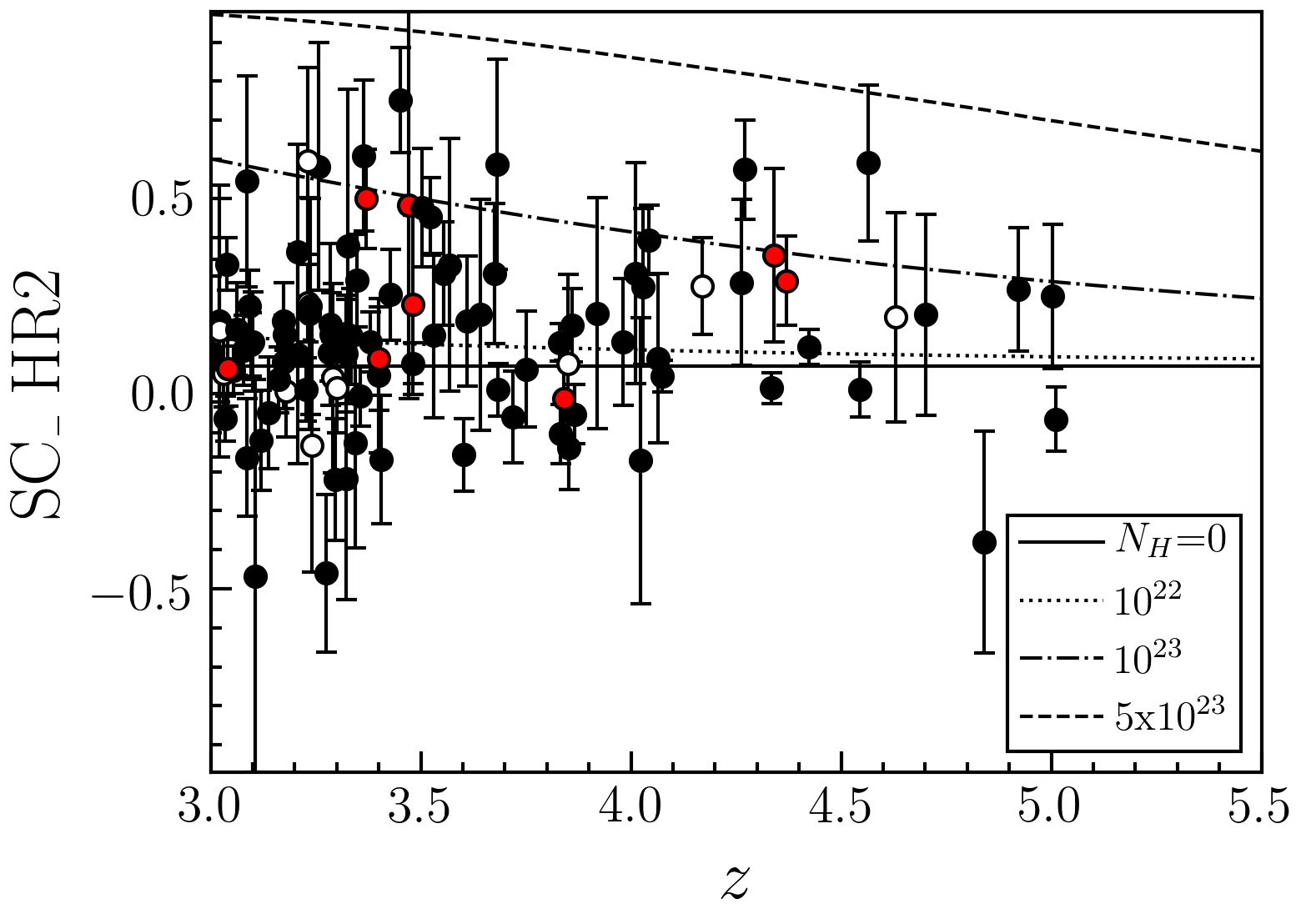}
\caption{ Distribution of \K16\ sources in X-ray hardness ratio
(defined via the 1–2~and~0.5–1~keV fluxes, see the
text) and redshift. The quasars with known $\zspec$ are indicated by the black symbols; the objects from the \K16\ catalog whose
spectroscopic redshifts were measured already after the publication of the catalog \citep{khorunzhev17,khorunzhev17b,khorunzhev18b}
are highlighted by the red color; the objects for which only $\zphot$ are known are designated by the open symbols. 
The lines indicate the dependence of the hardness ratio on redshift and intrinsic absorption column density for a 
power-law spectrum \mbox{with $\Gamma=1.8$.}}
\label{fig:hardredsh}
\end{figure}

Our sample consists of unabsorbed or weakly absorbed
X-ray quasars with an intrinsic absorption
column density $\NH<10^{23}$~cm$^{-2}$. 
This is evidenced by the distribution of sources in X-ray hardness ratio
(\3XMMDR4 data) and redshift presented in Fig.~\ref{fig:hardredsh}.
The hardness ratio \mbox{($SC_{HR2}=(H-S)/(H+S)$)} is
defined via the photon count rates in the 1–2~keV (H)
and 0.5–1~keV (S) bands. For comparison, Fig.~\ref{fig:hardredsh}
shows the redshift dependences of the hardness ratio
expected for a power-law spectrum with a slope $\Gamma=1.8$
and various absorption column densities. 
We see that only a few sources from the sample would have
\mbox{$\NH\simeq 10^{23}$~cm$^{-2}$.} 
The rest should have a lower intrinsic absorption.

The absence of heavily absorbed X-ray sources in the sample is related to the method of selecting
distant quasars in the optical band.
When compiling the \K16\ catalog based on shallow broadband photometry,
we selected type~1 quasars with an ultraviolet
excess (in the quasar’s rest frame) and an absorption
gain behind the Ly$\alpha$ line. The same selection effect is
described, for example, in \cite{kalfonzou14},
where the fraction of luminous quasars with absorption
$\NH>10^{23}$~cm$^{-2}$ was about 10\% due to a similar
selection method, and \cite{vito14}, where from
Table~1 it can be seen how the fraction of absorbed
quasars drops with decreasing sensitivity of X-ray and optical surveys.

\subsection{The Subsample of Fainter Quasars from the Paper by Vito et al.}

The quasar luminosity function has the form of a
power law with a break (\citealt{boyle88,miyaji00}; see Eq. (\ref{eq:2slope}) below). 
A sample spanning a wide luminosity range is needed to determine all
parameters of the luminosity function model. 
The region near the break in the luminosity function, where
the power-law slope changes, is especially important.
All objects in the K16 catalog have luminosities higher than the break luminosity 
($\LX210 \simeq 4\times 10^{44}$~erg/s; \citealt{vito14}). 
Therefore, it was decided to supplement the K16 list of luminous quasars
by the sample from \cite{vito14}, which contains
many objects near the break luminosity.

The catalog of X-ray quasars at $z>3$, based on
which we constructed the luminosity function, is presented
in the paper by \cite{vito14}. 
Almost all of the sources have spectroscopic measurements
or reliable estimates of the redshift obtained from
deep photometric survey data in medium-band filters.
Therefore, it is convenient to use the sample by 
\cite{vito14} to extend \K16\ to lower luminosities.
From the catalog \cite{vito14} it is easy to
extract the \V14U\ subsample of unabsorbed sources
(absorption column density $\NH\leq 10^{23}$~cm$^{-2}$) for a
better correspondence to the \K16 sample.

The original sample by \cite{vito14} consists
of 141 X-ray sources at redshifts \mbox{$3<z<5.1$} and
was obtained from the data of four deep \mbox{X-ray}
surveys: Chandra Deep Field South (CDFS, \citealt{xue11}, 
Chandra Cosmos Survey (\mbox{C-COSMOS}, \citealt{elvis09}), XMM-Newton Cosmos Survey
(\mbox{XMM-Cosmos}, \citealt{hasinger07}), and
\mbox{Subaru/XMM-Newton Deep Survey} (SXDS, \citealt{ueda08}). 
In these surveys the optical identification
completeness of X-ray sources is higher than 95\%.
The total area is 3.3~sq.~deg. 
A total of three sources
have 2-10~keV luminosities~$\geq10^{45}$~erg/s. 
Only one of them is unabsorbed.

The subsample of \Nmyvito \ unabsorbed sources (\V14U)
used in our paper consists of quasars with luminosities
$8\times10^{42}<\LX210<1.04\times10^{45}$~erg/s.
The source {ID 5120} was excluded from the XMM-COSMOS survey,
because it is a star \citep{lilli07}.
The most distant source {ID~2220} 
($\zspec=5.07$, $\LX210 =4.7\times 10^{44}$~erg/s)
was found in the C-COSMOS survey \citep{elvis09}. 
The most luminous source {ID~926} (\mbox{$\LX210 =1.04\times 10^{45}$~erg/s}, $\zspec=3.264$) was found 
in the SXDS survey \citep{ueda08}. In Fig.~\ref{fig:catflux} the X-ray fluxes, luminosities,
and redshifts of the sources from the \V14U\ subsample
are compared with the corresponding characteristics
of the sources from the \K16 subsample.

\section{THE SURVEY AREA}

\begin{figure}[!ht]
\centering
\includegraphics[width=\linewidth]{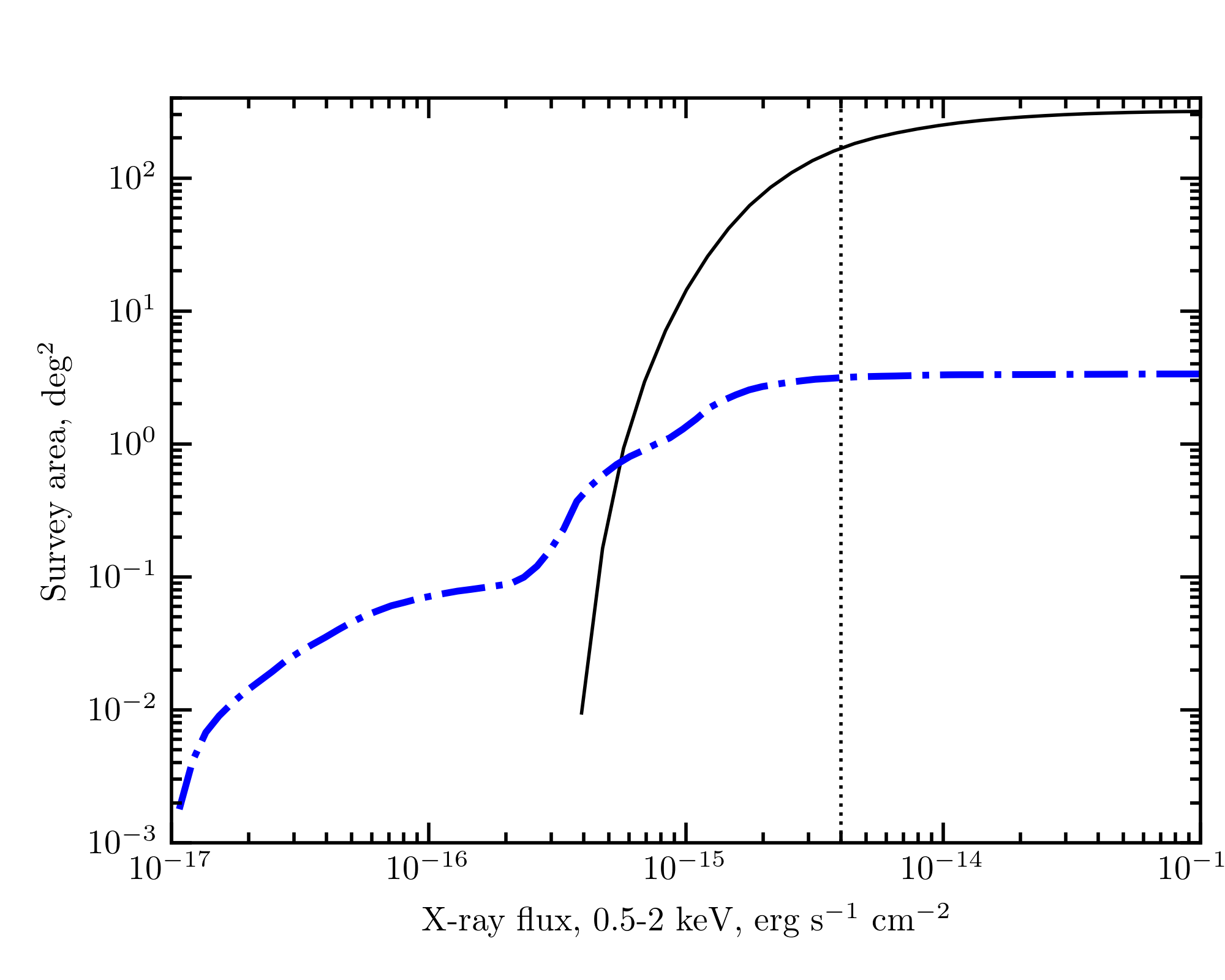}
\caption{The \3XMMDR4\ survey area (black solid line) and total area of deep surveys for unabsorbed sources (\citealt{vito14}, the blue dash-dotted line). 
The vertical dotted line marks the lower boundary the 0.5--2~keV X-ray flux
$4\times 10^{-15}$~erg/s/cm$^2$  for the \K16\ subsample of objects with $\LX210>10^{45}$~erg/s.} 
\label{fig:Area} 
\end{figure}

To calculate the space density of sources, we need
to know how the sky coverage area of the X-ray survey
changes with sensitivity. For the V14U subsample
of unabsorbed low-luminosity sources we took the
corresponding area for unabsorbed sources from \cite{vito14} (see Fig.~\ref{fig:Area}).

To calculate the area of the \XMM \ serendipitous
survey, we selected the pointings (OBSID) that
were used to construct the \mbox{\3XMMDR4}\footnote{xmmssc-www.star.le.ac.uk/Catalogue/3XMM-DR4/} catalog
of X-ray sources \citep{watson09} and were
previously used by us \citep{khorunzhev16} to roughly estimate the survey area: the sources
must be at Galactic latitudes $|b|>20^\circ$ and fall into
the SDSS region. Using the utility task \textit{esensmap}  of the \textit{XMM-Newton Science Analysis System}, 
we constructed the sensitivity maps of individual pointings
(in counts/s/PSF) for the detection threshold
$\detML>6$ in the range 0.2-12~keV for the total
exposure of all the detectors involved in this pointing.
The original \3XMMDR4\ catalog of X-ray sources
was compiled precisely with this detection threshold
($\detML>6$). In the case where the nearby pointings
overlapped to form a mosaic, we chose fields from
the pointing with the best sensitivity to construct the
sensitivity map in the overlapping region.

When the \3XMMDR4\ catalog of sources was
compiled, the counts from all the operating detectors
for an individual pointing were taken into account.
Each mode of operation of the \XMM\ detectors
is characterized by its count rate-to-flux conversion
factor\footnote{heasarc.gsfc.nasa.gov/w3browse/all/xmmssc.html}. 
To convert the sensitivity map from
counts/s to erg/s/cm$^2$ (the 0.2-12~keV band),
we calculated the effective conversion factor from the
following formula:

\begin{equation}
\begin{aligned}
\langle ECF\rangle=\dfrac{\sum\limits_{i=1}^{\Ndet} EXP_{i} \times ECF_{i}}{\langle EXP\rangle} , \\
\langle EXP\rangle=\frac{1}{\Ndet}\sum\limits_{i=1}^{\Ndet}EXP_{i},
\end{aligned}
\label{eq:ECF}
\end{equation}
where $\Ndet$ --- is the number of operating detectors in a given pointing, 
$EXP_{i}$ is the exposure map of the \textit{i}-th detector, 
$ECF_{i}$ is the count rate-to-flux conversion factor 
(the 0.2--12~keV band) for the mode of operation
of $i$-th detector, $\langle EXP\rangle$ is the mean exposure time. 
The sensitivity map was then divided
by the map of the effective count rate-to-flux conversion
factor $\langle ECF\rangle$. The fluxes were converted
from the 0.2-12~keV band to the 0.5-2~keV band
of interest to us by assuming a power-law spectrum
of sources with a slope \mbox{$\Gamma=1.7$} and absorption 
$N_H=10^{20}$~cm$^{-2}$ (roughly corresponding to the absorption
in the Galactic interstellar medium). We used precisely
$\Gamma=1.7$, because the tabulated count rate-to-flux
conversion factors for the \XMM\ bands
are given for this slope. For all of the chosen fields
we then obtained the cumulative number distribution
of pixels (with a flux below the specified one) and
constructed the dependence of the survey area on X-ray
flux (see Fig.~\ref{fig:Area}).

The total area of the overlap between \3XMMDR4\
and \SDSS\ is 320~sq.~deg, which is almost a factor
of 100 larger than the total area of the \V14U\ survey
from \cite{vito14}.

\section{THE SDSS IDENTIFICATION COMPLETENESS OF X-RAY SOURCES}
\begin{figure}[!ht]
\centering
\includegraphics[width=\linewidth]{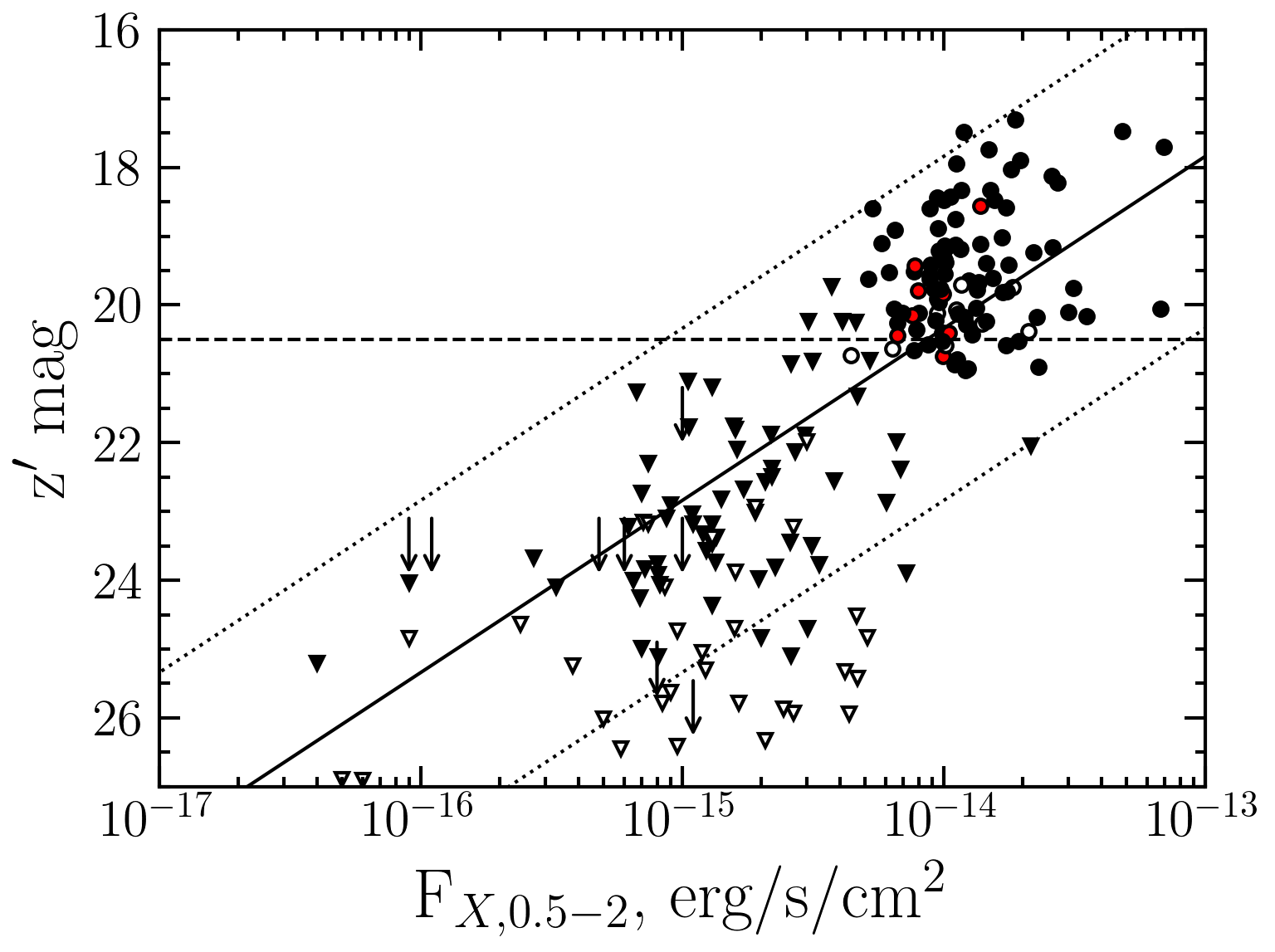}

\caption{Apparent $\zmag$ magnitudes and X-ray (0.5-2~keV) fluxes of the sources from the \K16 (circles) and \V14U (triangles) subsamples. 
The filled symbols indicate the objects with spectroscopic redshifts ($\zspec$); the objects from the \K16 catalog
for which the spectroscopic redshifts were measured already after the publication of the catalog \citep{khorunzhev17,khorunzhev17b,khorunzhev18b} are red highlighted. The open symbols indicate the objects for which there are only photometric
redshift estimates ($\zphot$). The arrows mark the lower limits on $\zmag$. The solid line indicates the ratio of the \mbox{X-ray} and optical fluxes \mbox{$f_X/f_O=1$}; the dotted lines indicate $f_X/f_O=0.1$ and $f_X/f_O=10$. 
The horizontal dashed line indicates the effective threshold of the K16 catalog $\zmag=20.5$.}
\label{fig:magflux}
\end{figure}

To obtain trustworthy photometric reshifts $\zphot$ when constructing
the \K16\ catalog \citep{khorunzhev16},
we used only reliable optical sources, with an error
of the apparent magnitude \mbox{$\Delta \zmag <0.2$} in the \SDSS\  $\zmag$
band, corresponding to an effective detection threshold \mbox{$\zmag\simeq 20.5$}. 
Fainter sources
were not included in the \K16\ catalog. This could
skew the sample toward optically luminous quasars
(see Fig.~\ref{fig:magflux}). This figure shows the distribution of
type~1 quasars from the \K16\ sample in X-ray flux and
apparent magnitude in the \SDSS \ $\zmag$ band.

For comparison, Fig.~\ref{fig:magflux} shows the unabsorbed
sources from the \V14U subsample that has an almost
100\% optical identification completeness. The
apparent magnitudes of the X-ray sources in the $\zmag$
band were taken from \cite{civano12,capak07} for C-COSMOS and XMM-COSMOS respectively, 
and \cite{akiyama15} for SXDS. 
The magnitudes from the GOODS \citep{giavalisco04} and
GEMS \citep{caldwell08} photometric surveys in
the z850 filter, whose range roughly coincides with
the \SDSS \ $\zmag$ band, were used for the CDFS survey \citep{xue11}.

\begin{figure}[!ht]
\centering
\includegraphics[width=\linewidth]{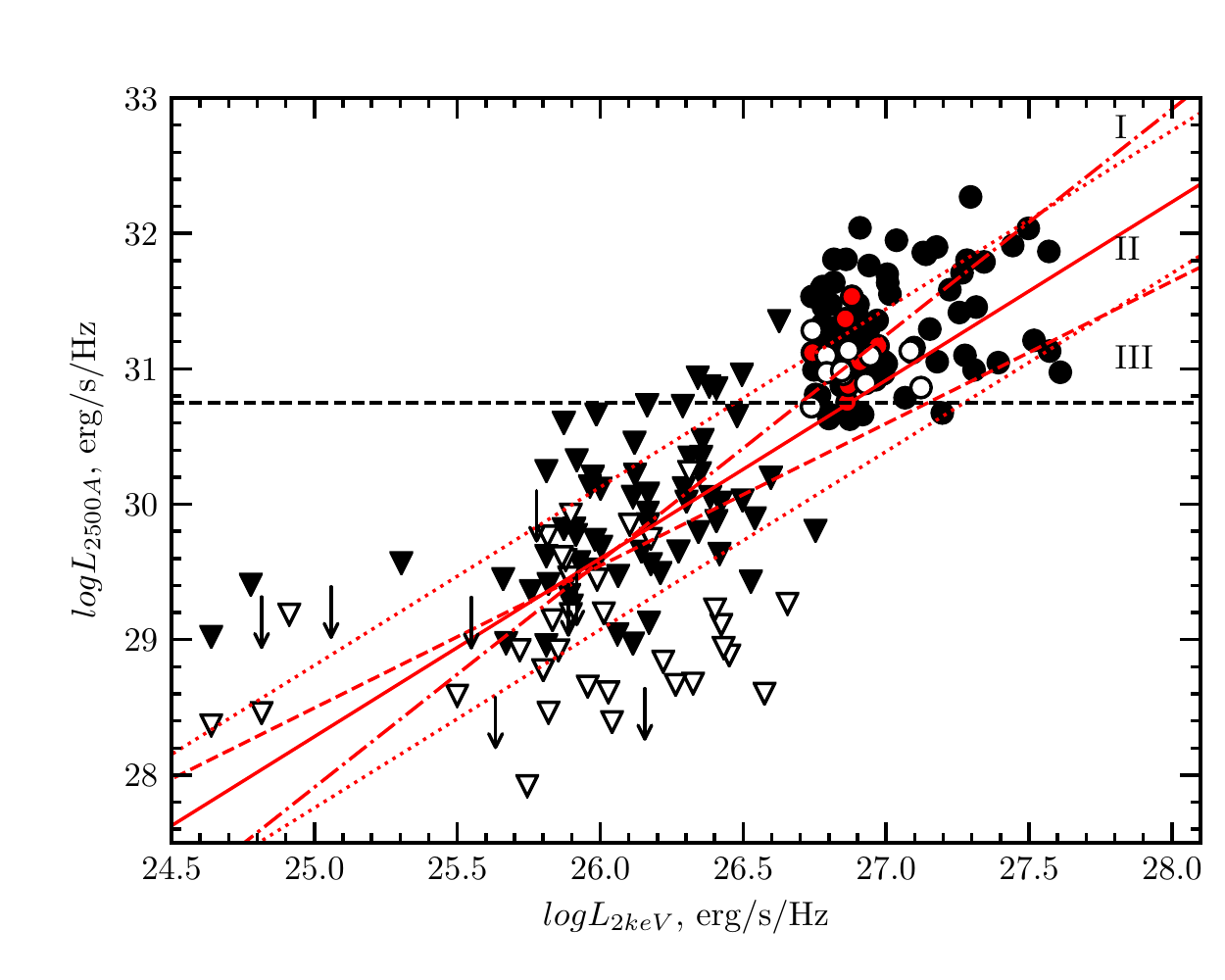}
\caption{
Relation between monochromatic optical luminosity $\l2500A$ at wavelength 2500~\AA\ and monochromatic
X-ray luminosity $\kev2l$ at energy 2~keV for the \K16\ and \V14U \ subsamples. 
The designations are the same as those in Fig.~\ref{fig:magflux}.
The red solid line labeled by "II"\ indicates the \LXLO\ relation from \cite{lusso10}, which was taken
as a basis in calculating the correction for incompleteness. 
The red dash–dotted and dashed lines labeled by "I"\ and "III"\ indicate other \LXLO\ relations from \cite{lusso10} that are used as the minimum and maximum corrections
for incompleteness. The dotted lines indicate the $\pm0.4$ scatter for the \LXLO\ relation II. 
The horizontal dashed line
indicates the threshold luminosity $\l2500A$ that a source with an apparent magnitude $\zmag = 20.5$ at redshift $z=3$ whose spectrum is described by the template from \cite{vandenberk01} will have.
}
\label{fig:monochromatic}
\end{figure}

It can be seen from Fig.~\ref{fig:magflux} that the X-ray-to-optical
flux ratio for most of the \K16\ sources is less
than unity ($f_X/f_O<1$), while most of the objects
from the \V14U\ sample have \mbox{$f_X/f_O>1$}. This can
be related in part to the known nonlinear correlation
between the optical and X-ray luminosities of
quasars: the higher the bolometric luminosity of an
object, the smaller the ratio $f_X/f_O$ (see, e.g., \citealt{lusso10,lusso17})\footnote{However, there is evidence in a number of papers that the dependence of $f_X/f_O$ can be approximately linear \citep{sazonov12,marchese12}.} The fact that the threshold \mbox{$\zmag \simeq 20.5$} used in constructing the \K16 \ catalog turns out to
be insufficient for the detection of all high-luminosity
X-ray quasars at $z>3$ is apparently more important.

\subsection{The Method of Calculating the Correction for Incompleteness}

The completeness of quasars in the \K16\ subsample
cannot be estimated using the observational data
of X-ray surveys with an area of $\sim 20$~sq.~deg, for
example, \XMMXXL. The size of such surveys is too
small to detect a sufficient number of distant quasars
with luminosities $\geq 10^{45}$~erg/s. 
Therefore, we used
the relation between the X-ray,  $\kev2l$, and optical,
 $\l2500A$, monochromatic luminosities of type~1
quasars \citep{lusso10,marchese12} 
to determine the missed fraction of X-ray quasars with
an apparent magnitude $\zmag > 20.5$.

\cite{lusso10} used a subsample of type~1
quasars from the deep {XMM-COSMOS} survey to
investigate the \LXLO \ relation. In this sample
60\% of the sources have spectroscopically confirmed
redshifts. Most of the quasars at $z>3$ from
the sample by \citep{lusso10} are present in the
\V14U sample. Subsequently, \cite{marchese12}
obtained similar results for a spectroscopically complete
sample of optically luminous quasars selected in
the X-ray and ultraviolet bands.

We considered three variants of the \LXLO \ relation: 
\begin{align*}
\mbox{I}&: \log \l2500A =1.669 \log \kev2l - 13.815 , \\
\mbox{II}&: \log \l2500A =1.316 \log \kev2l - 4.616 , \\
\mbox{III}&: \log \l2500A =1.050 \log \kev2l + 2.246  . 
\end{align*}
These relations were taken from \citep{lusso10}:
I --- when $\l2500A$ was used as an independent variable,
III --- when $\kev2l$ was used as an independent
variable (see also \citealt{marchese12}), II --- the
bisector between relations I and III. The scatter of
individual measurements about relation II is characterized
by a dispersion of 0.37 \citep{lusso10}.
Using a sample of unabsorbed quasars from the \XMMXXL \
survey as an example, \citep{georgakakis15}
showed that \LXLO \ agrees with relation II
with a dispersion of~0.4.

Following algorithm to calculate
the correction for K16 subsample incompleteness had been applied.
Assuming a power-law X-ray spectrum with a photon
index $\Gamma$=1.8, we calculated the monochromatic
luminosity $\kev2l$ at energy 2~keV via the
X-ray luminosity $\LX210$ in the quasar’s rest frame.
Next, we determined its optical monochromatic luminosity
$\l2500A$ at wavelength 2500~\AA \ via relations \mbox{I---III}. 
The monochromatic luminosity $\l2500A$ was then
converted to the apparent magnitude in the \SDSS \ $\zmag$
band in the observer’s frame using a template of the
quasar spectrum \citep{vandenberk01}. As
a result, we obtain the mean expected value of $\zmag$ for
a quasar with an X-ray luminosity $\LX210$. 
Finally, by assuming that the \LXLO\ scatter is
Gaussian and has a dispersion $\sigma$=0.4, we calculated
the probability that the quasar would be brighter
than $\zmag$=20.5. The probability is the correction for incompleteness
$\Theta(L,z)$ describing the fraction of quasars with
a luminosity $\LX210$ that are optically brighter than $\zmag \leq 20.5$. 
For the \V14U\ subsample we assumed that $\Theta(L,z)=1$.

Relation II was used to calculate the main correction
for incompleteness, while relations I and III were
used as the minimum and maximum corrections, respectively.
Thus, relations I and III are assumed to
limit the possible systematic scatter of the correction
for incompleteness. Figure~\ref{fig:monochromatic} shows relations I, II,
and III. The scatter plot between the luminosity $\kev2l$
calculated via $\LX210$ for $\Gamma$=1.8 and the luminosity
$\l2500A$ derived via the measured $\zmag$ using the template
from \cite{vandenberk01} is also shown there for
the sources from the \K16\ and \V14U\ subsamples. The
derived corrections for incompleteness are used below
to calculate the quasar luminosity function.

\section{THE LUMINOSITY FUNCTION}

Below by the X-ray luminosity function $\phi(\LX210,z)$
we understand the number density of quasars per unit
interval of the decimal logarithm of the X-ray luminosity
(in the 2-10~keV band in the quasar’s
rest frame) as a function of luminosity and redshift.
We investigated the luminosity function by both
parametric and binned (nonparametric) methods.

\subsection{Analytical Estimates of the Luminosity Function}

We considered several popular empirical X-ray luminosity
function models for AGNs. As their basis
is regarded a smoothly-connected two power-law form with a break \citep{piccinotti82,boyle88,miyaji00}:
\begin{equation}
\phi=\frac{A}{\big(\frac{L_{\rm X}}{L_*}\big)^{\gamma_1}+
  \big(\frac{L_{\rm X}}{L_*}\big)^{\gamma_2}}, 
\label{eq:2slope}
\end{equation}
where А is the normalisation, $L_*$ is the break luminosity,
$\gamma_1$ and $\gamma_2$ are the slopes of the function below and above the break luminosity,
 $L_{\rm X}$ is the \mbox{X-ray} luminosity.
 In all of the models listed below we
assume a reference redshift parameter $\zmin=3.0$, see also \cite{vito14}.

To obtain the Pure Luminosity Evolution (PLE)
model \citep{longair70}, the break luminosity
$L_*$ in Eq. (\ref{eq:2slope}) needs to be multiplied by 
\begin{equation*}
\elum(z) = [(1+z)/(1+\zmin)]^{\plum} ,
\end{equation*}
where $\plum$ is the luminosity evolution factor. 
It is assumed in the model that the total density of quasars
does not change with time, but the shape of the density
dependence, the ratio of bright and faint sources,
changes. \cite{vito14} showed that the PLE
model is poorly suited to describing the distribution
of quasars at high redshifts.

To obtain the Pure Density Evolution (PDE)
model \citep{schmidt68}, the normalisation A in
Eq.~(\ref{eq:2slope}) needs to be multiplied by
\begin{equation*}
\eden(z) = [(1+z)/(1+\zmin)]^{\pden} ,
\end{equation*}
where $\pden$ is the density evolution factor. In the
PDE model it is assumed that the density of sources
changes with time, while the ratio of the densities of
bright and faint quasars is retained.

The more complex Independent Luminosity and
Density Evolution (ILDE) model \citep{yencho09}
is obtained from the PDE model by multiplying the
break luminosity $L_*$ by
\begin{equation*}
\elum(z) = [(1+z)/(1+\zmin)]^{\plum}.
\end{equation*}  
The Luminosity and Density Evolution (LADE)
model \citep{aird10} is also considered. This
model is obtained from Eq. (\ref{eq:2slope}) by multiplying the break luminosity $L_*$ by
\begin{equation*}
\elum(z) = [(1+z)/(1+\zmin)]^{p_{lum}} ,
\end{equation*} 
and multiplying the normalization $A$ by
\begin{equation*}
\eden(z) = 10^{\pden(z-\zmin)}.
\end{equation*} 
In this model an exponential time dependence of the quasar
density is assumed, in contrast to a powerlaw
dependence in PDE and ILDE. The last Luminosity
Dependent Density Evolution (LDDE) model
\citep{schmidtgreen83} under consideration is obtained
by multiplying $A$ in Eq. (\ref{eq:2slope}) by
\begin{equation*}
\eden(z) = [(1+z)/(1+\zmin)]^{\pden+\beta(\log L-44)}, 
\end{equation*} 
where $\beta$ is an additional parameter that accounts for the luminosity 
dependency.
The original LADE and LDDE models contain much more parameters,
because they were constructed to describe
large data sets in a wide range of redshifts (0.001–5)
and luminosities. Since the parameters degenerate at
high redshifts $z>3$, the LADE and LDDE models were simplified by \cite{vito14}.

To determine all parameters of the listed models,
we need samples spanning a wide luminosity
range $10^{43} \lesssim
\LX210\lesssim 10^{46}$~erg/s. The luminosities
of the \K16\ subsample objects exceed the break
luminosity $L_* \simeq 5\times10^{44}$~erg/s \citep{vito14}.
Consequently, by adding new sources at luminosities
$\LX210>10^{45}$~erg/s, we can improve significantly
the constraints only for some of the parameters (the
normalization and slope of the spectrum after the
break $\gamma_2$).

We used the maximum likelihood method to find
the best model. More specifically, using the \textit{scipy}\footnote{\url{http://www.scipy.org/}}~\textit{optimize} library, 
we minimized the following function:
\begin{equation}
\begin{aligned}
\Lagr(\theta) = \Lagr(\theta,\Omega_{K16},N_{K16})+ \\
+\Lagr(\theta,\Omega_{V14U},N_{V14U}),
\end{aligned}
\label{eq:likelihood1}
\end{equation}
where $\theta$ are set of the model parameters ($\theta$=[$\theta_1,\theta_2,...,\theta_k$]), terms
$\Lagr(\theta,\Omega_{K16},N_{K16})$ and $\Lagr(\theta,\Omega_{V14U},N_{V14U})$ are the
likelihood functions for the \K16\ and \V14U subsamples
described by Eq. (\ref{eq:likelihood}) given below, $\Omega_{K16}$ and
$\Omega_{V14U}$ are the dependences of the survey's coverage area on sensitivity presented
in Fig.~\ref{eq:likelihood}, $N_{K16}$ and $N_{V14U}$ are the
object counts in the corresponding subsamples. According
to Fig.~\ref{fig:Area}, the coverage area $\Omega = \Omega(\FX0.52(L,z)) = \Omega(L,z)$ is determined respect to the flux $\FX0.52$ that
is expected from a source with a photon index of
the X-ray spectrum $\Gamma=1.8$, luminosity $\LX210$, and
redshift $z$. The correction for incompleteness $\Omega_{K16}=\Omega_{K16}(z,L)\times\Theta(L,z)$ is included into the dependence of
the area $\Omega_{K16}$ for the \K16\ sample.

The likelihood function for each subsample in Eq.~(\ref{eq:likelihood1}) is written as
\begin{equation} 
\begin{aligned}
\Lagr(\theta,\Omega,N) = -2 \sum^{N}_{i=1}\ln[\int\phi(L,z_{i},\theta) p(d_i|L) d\log L]+ \\
+2\iint\phi(L,z,\theta)\Omega(L,z)\frac{dV}{dz}d\log L\,dz,
\end{aligned}
\label{eq:likelihood}
\end{equation}
where $\phi(L,z,\theta)$ is the X-ray luminosity function
model, $z_{i}$ is the redshift $i$-th source, $N$ is the total subsample source counts, 
$dV/dz$ is the differential
comoving volume per unit sky area, and $p(d_i|L)$ is
the probability density to detect a source with a data
set $d_i$ provided that its luminosity is $L$. The double
integral in Eq. (\ref{eq:likelihood}) is taken in the redshift interval
$3<z<5.1$ and the following luminosity ranges:
$42.85 < \log
L<45.3$ for the \V14U\ subsample and
$45.0 < \log L<45.9$ for the \K16 subsample.

For the \K16\ sources the data set $d_{i}$ of the function
$p(d_{i}|L)$ includes: the expected number of counts
$s=s(L)$ that depends on luminosity; the number of
recorded source ($s_0$) and background ($b_0$) counts in
the 0.2-12~keV band. The quantity $p(d_{i}|L)$ itself
expresses the probability to record the total number
of counts $(s_0+b_0)$ from $i$-th source:
\begin{equation} 
p(d_{i}|L)=\frac{(s+b_0)^{(s_0+b_0)}}{(s_0+b_0)!}e^{-(s+b_0)}.
\label{eq:eddshift}
\end{equation}
This approach takes into account the Poissonian nature
of the detection of photons and the related Eddington
bias of the X-ray luminosity function \citep{georgakakis08,aird10}.

When calculating the expected number of counts ($s$) in the 0.2-12~keV 
band from a source with luminosity
$\LX210$ at redshift $z_{i}$, we assumed a powerlaw
X-ray spectrum with a photon index $\Gamma$=1.8. For
each source we used the count rate-to-flux conversion
factor calculated as the ratio of the 0.2-12~keV
count rate {EP\_8\_Rate} to the corresponding X-ray
flux {EP\_8\_Flux}. The values of $s_0$, {EP\_8\_Rate} and
{EP\_8\_Flux} are taken from the \3XMMDR4 catalog.

The \3XMMDR4\ catalog provides only the averaged
density of background counts per pixel of the
background map in a set of energy bands, while the
counts from the source and the count rates are given
with the background subtracted. Consequently, it
is impossible to accurately reconstruct the number
of background counts $b_0$ from the \3XMMDR4\ data.
We checked that the background for most of the
\K16\ sources made a minor contribution to the total
number of counts, i.e., $b_0$ accounts for a few percent
of $s_0$. Therefore, in Eq. (\ref{eq:eddshift}) we neglect the background
counts and assume $b_0=0$.

For the \V14U\ subsample no correction is made
for the Eddington bias. Therefore, for $i$-th source  from
the \V14U\ subsample $p(d_i|L)$ is a delta function of the
observed luminosity $L_{i}$.

The 1$\sigma$ confidence intervals are computed
by varying each  $i$-th parameter $\theta_i$ in the vicinity of its best value
The boundary limits $\theta_{i,{\rm min}}$ and $\theta_{i,{\rm max}}$
for which the value of the likelihood function differed
from its value at the minimum by one ($\Delta \Lagr=1$) defines the confidence interval. 
At the same time, the remaining parameters are left free.

In this way we fitted the data of the joint sample
of unabsorbed \V14U and \K16\ quasars by the PDE,
PLE, ILDE, LADE, and LDDE models. Our estimates
of the parameters and their statistical errors for
the incompleteness correction II are given in Table~\ref{tab:LFA1}.
The parameter estimates for corrections I and III are also given there in parentheses.

All of the luminosity function models reproduce
accurately the observed number of sources (\Nmylumtot)
when integrated over the $\log L$--$z$ space. However,
the calculated number of sources in the \V14U\ sample
is overestimated relative to their true number and,
accordingly, the calculated number of sources in
the \K16 sample is underestimated; the higher the
degree of incompleteness correction, the greater the
difference between the calculated numbers of sources
from \V14U\ and \K16. Nevertheless, these deviations
remain within 1$\sigma$ according to a Poisson statistic of source counts for corrections
I and II and 2$\sigma$ for correction III. An increase in the
calculated number of sources in the V14U subsample
with a high completeness suggests that correction III
may be excessive.

We used the Akaike information criterion (AIC,
\citealt{akaike74}) and the Bayesian information criterion
(BIC, \citealt{schwarz78}) to determine the best luminosity
function model from the set being investigated (see, \cite{fotopoulou16,sazonovkhabibullin17}). 
For a logarithmic likelihood
function (see Eq.~(\ref{eq:likelihood})) the formula for AIC is \mbox{$AIC=2k+\Lagr$},
where k is the number of model parameters, $\Lagr$
is the value at the minimum of the likelihood function.
BIC was calculated from the formula $BIC=k\,\ln n+\Lagr$, 
where n is the number of objects in the observational
sample. BIC is a modification of AIC and
is better suited to comparing models with different
numbers of parameters. The best model will have the
lowest AIC and BIC values.

For each model we obtained the differences $\Delta AIC = AIC-AIC_{0}$ and $\Delta BIC = BIC - BIC_{0}$, where $AIC_{0}$ and $BIC_{0}$ are the values of the criteria for
the best model. The larger the value of $\Delta AIC$ and
$\Delta BIC$, the lower the probability that a given model
is suitable for describing the data.

The derived values of $\Delta AIC$ and $\Delta BIC$ for the
incompleteness correction II are given in Table~\ref{tab:LFA1}.
LDDE turns out to be the best model. The deviations
$|\Delta BIC|<6$ may be deemed statistically insignificant.
Consequently, the LDDE, LADE, ILDE and
PDE models are equally suitable for describing the
data. Only the PLE model has $|\Delta BIC|\gg 6$ and
reproduces the observational data more poorly than
do the remaining ones. Therefore, it may be excluded
as untenable.

Thus, the set of admissible X-ray luminosity function
models for distant type~1 quasars turned out to
be the same as that in a number of previous papers,
where samples of quasars including absorbed objects
were used \citep{vito14,georgakakis15,ranalli16}.

\begin{figure*}[pth!]
\label{fig:aodlf}
\begin{minipage}{\textwidth}
\includegraphics[width=.48\textwidth]{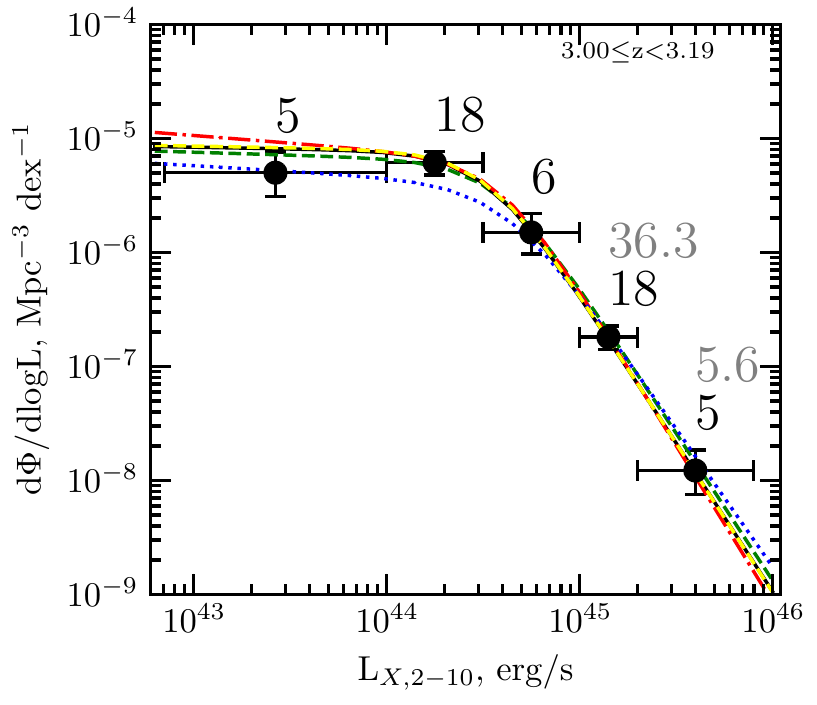}
\includegraphics[width=.48\textwidth]{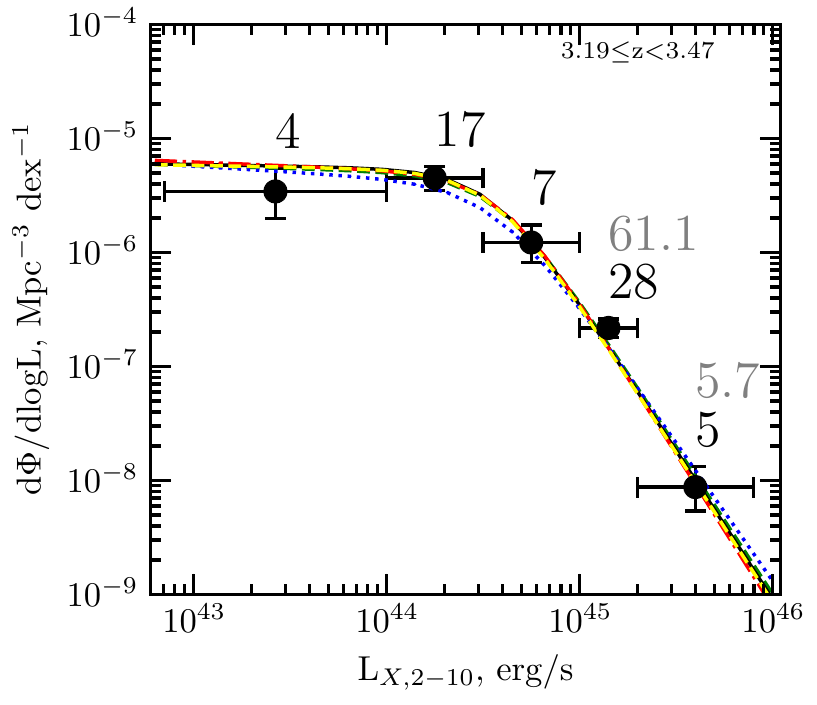}
\end{minipage}
\hspace{\fill}
\begin{minipage}{\textwidth}
\includegraphics[width=.48\textwidth]{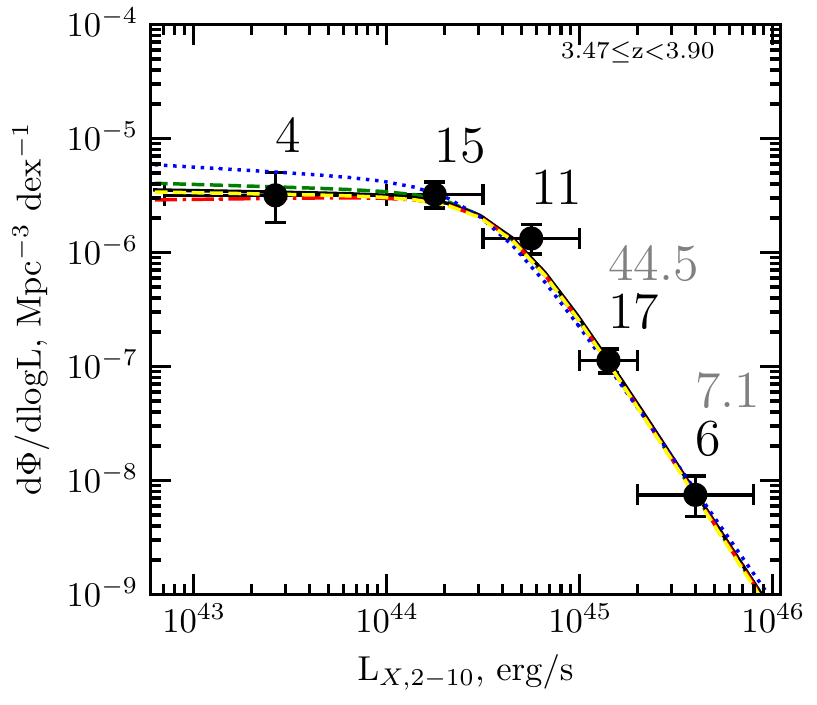}
\includegraphics[width=.48\textwidth]{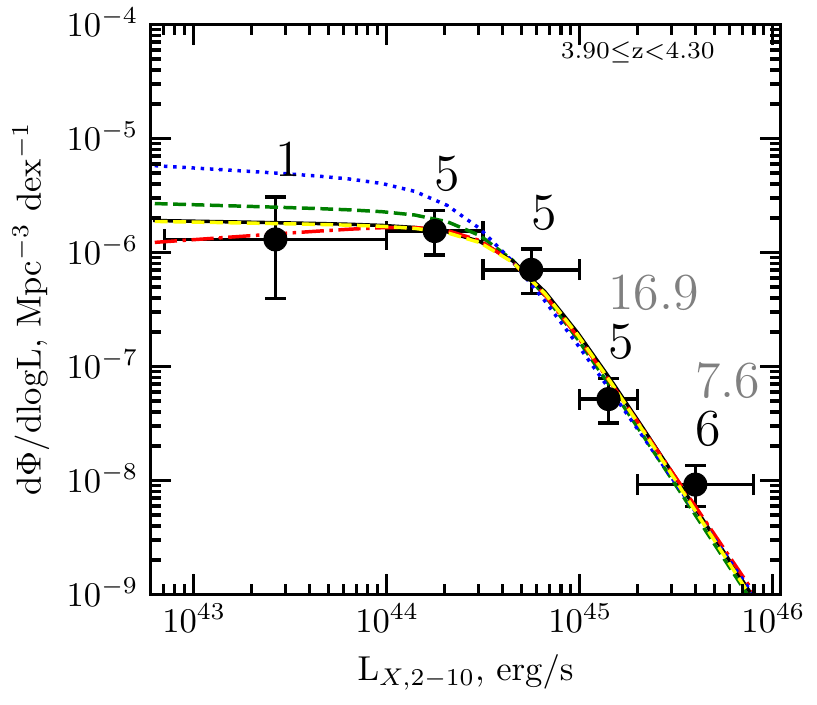}
\end{minipage}
\hspace{\fill}
\begin{minipage}{\textwidth}
\includegraphics[width=.48\textwidth]{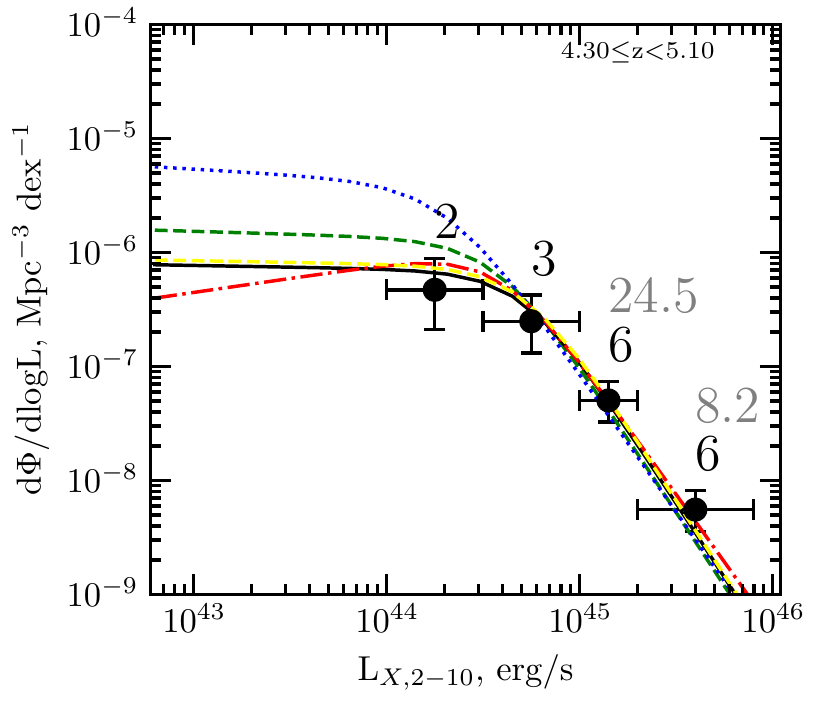}\hfill
\end{minipage}
\begin{minipage}{\textwidth}
\caption{
The differential luminosity function of type~1 quasars as a function of redshift based on the \K16\ and
\V14U \ subsamples. The points with error bars indicate the binned (nonparametric) estimate of the luminosity function obtained by the
$1/\Vmax$ method using the incompleteness correction II. The black font indicates the number of sample objects in the $\Delta\log L$--$\Delta z$
bins under consideration, while the gray font indicates the expected total number of objects corrected for the \K16 subsample
incompleteness. The black solid, red dash–dotted, green dashed, yellow dashed, and blue dotted lines represent the LADE,
LDDE, PDE, ILDE and PLE models, respectively. 
}
\end{minipage}
\end{figure*}

\subsection{The binned Luminosity Function ($1/\Vmax$)}

A nonparametric estimate of the X-ray luminosity
function is an estimate of the space density of
quasars calculated separately for each of the specified
$\Delta\log L$--$\Delta z$ bins from the sample objects falling into
these bins. We performed such a calculation by the
method described in \citep{georgakakis15}. The
space $42.85<\log L<45.9$, $3<z<5.1$ was divided
into $\Delta \log L$--$\Delta z$ bins close to those used in \cite{vito14}. 
The binning scheme and the number
of sources in the corresponding bins are shown in
Table~\ref{tab:bindistr} and Fig.~\ref{fig:catflux}.

Assuming that within each $\Delta\log L$--$\Delta z$ bin 
(which contains $N$ sources) the luminosity function is constant,
i.e., $\phi(L,z,\theta)= C$, we can search for $C$ by the
maximum likelihood method using Eqs. (\ref{eq:likelihood1}) and (\ref{eq:likelihood}),
where the constant $C$ in Eq. (\ref{eq:likelihood}) is the only parameter:

\begin{equation} 
\begin{aligned}
\Lagr(C,\Omega,N) = -2 \sum^{N}_{i=1}\ln[C p(d_i|L) d\log L]+ \\
+2\iint C\Omega(L,z)\frac{dV}{dz}d\log L\,dz.
\end{aligned}
\label{eq:vmax}
\end{equation}

It is easy to show that this function has a minimum at
\begin{equation} 
\begin{aligned}
C = \frac{N}{\iint \Omega(L,z)\frac{dV}{dz}d\log L\,dz}, 
\end{aligned}
\label{eq:vmaxpagecarrera}
\end{equation}
which closely corresponds to the expression from \cite{marshall83,pagecarrera00}
for calculating the luminosity function by the $1/V_{max}$ method.

The nonparametric estimate of the luminosity
function with the incompleteness correction for the
K16 subsample obtained in this way is presented in
Fig.~6. We see that the analytical luminosity function
models pass well through the points obtained by the
$1/V_{\rm max}$ method, with the points based on the K16
sample lying on the extension of the law of powerlaw
decline in the density of quasars at luminosity
$\LX210>10^{45}$~erg/s. It became possible to obtain
significant density estimates for distant quasars of
such high luminosities only by using the sensitive
\XMM\ X-ray survey with a large area
($\sim170$~sq.~deg at a flux \mbox{$\sim4\times10^{-15}$~erg/s/cm$^2$} for the \K16 \ sample).

It should be noted that the nonparametric estimate
of the luminosity function disregards the Eddington
bias, in contrast to the parametric estimate. Good
mutual agreement of both results suggests that the
Eddington bias in this case turns out to be insignificant
compared to the uncertainties associated with
the relatively small \K16\ sample size and the incompleteness
correction.

\section{EVOLUTION OF THE SPACE DENSITY OF DISTANT QUASARS}

\begin{figure*}[!ht]
\centering
\includegraphics[width=0.9\linewidth]{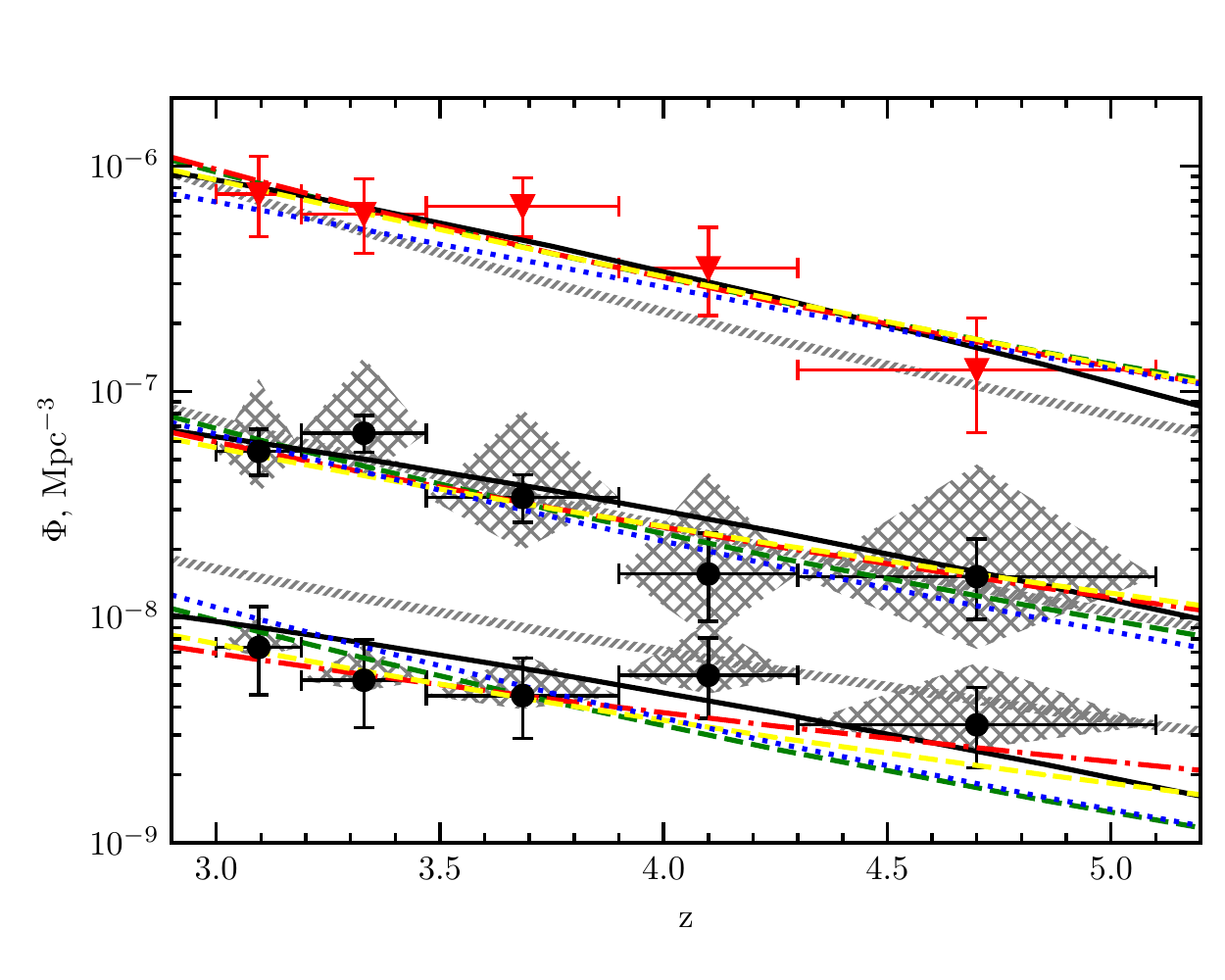}
\caption{
Evolution of the space density (in comoving coordinates) of distant luminous type~1 quasars. The
triangles indicate the density of quasars with $44.5\leq\LX210<45.0$ derived from the \V14U\ subsample; the circles indicate
the density of quasars with $45.0\leq\log\LX210<45.3$ and $45.3\leq\log\LX210<45.9$ derived from the \K16 \ subsample for the
incompleteness correction II. The hatched regions near the \K16 \ points are the scatter of densities related to the uncertainty
in the completeness of this subsample. The lines indicate the luminosity function models for the incompleteness correction II:
LADE (black solid line), LDDE (red dash-dotted line), ILDE (yellow dashed line), PDE (green dashed line), and PLE (blue
dotted line). The gray dashed line indicates the LADE model from \cite{georgakakis15}.
}
\label{fig:densredshift}
\end{figure*}

Using the above nonparametric estimate of the luminosity
function, let us consider the evolution of the
space density of high-luminosity quasars with redshift
in more detail luminosity bins: $44.5\leq\log\LX210<45.0$ (based
on the \V14U \ subsample) and $45.0\leq\log\LX210<45.3$, $45.3\leq\log\LX210<45.9$ (based on the \K16 \
subsample with the addition of one source from \V14U), see Fig.~\ref{fig:densredshift}. As expected, the \K16\ survey
has allowed reliable estimates of the space density
of luminous quasars ($\LX210>10^{45}$~erg/s) at high
redshifts to be obtained for the first time. The figure
also shows the various luminosity function models
discussed in this paper.

It can be seen from Fig. 7 that the comoving density
of luminous quasars $45.3\leq\log\LX210<45.9$)
changes by no more than a factor of 2 between $z=3$ and
$z=5$, while the density of lower-luminosity quasars
($44.5\leq\log\LX210<45.0$) decreases by an order of
magnitude (see also Vito et al. 2014; Kalfountzou
et al. 2014). In previous papers there has already
been evidence for slower evolution of more powerful quasars; 
now this tendency has become quite obvious
owing to the addition of the K16 subsample of luminous
quasars. Note that inaccurate knowledge of the
\K16 sample completeness introduces the main uncertainty
in our estimates of the density of luminous
quasars at luminosities log $\log\LX210
\approx 45$. However,
when the luminosity doubles ($\log\LX210 \geq
45.3$),
the density of sources drops by almost a factor of
10, the sources in the sample become fewer, and the
Poissonian errors become large than the scatter of
estimates related to incompleteness.

On the whole, the derived redshift dependence
of the quasar space density is consistent with the
density estimates by \cite{kalfonzou14} for unabsorbed quasars
at $z>3$. \cite{kalfonzou14} made an additional
selection (by photometric redshift for objects with apparent magnitude brighter
$\mbox{i}^\prime<21$) of distant quasars at $z>3$ and
for the first time estimated their space density at
luminosities $\log\LX210>44.7$ for a survey with an
area $\simeq 33$~sq.~deg. Quasar candidates selected by
$\zphot$ constitute half of the sample by \citep{kalfonzou14}.
We were able to improve significantly
the constraints on the density of very luminous
($\log\LX210>45.0$) quasars through an almost tenfold
increase in the sky coverage area compared to
\citep{kalfonzou14}. 
While in the \K16\ subsample 90\% of quasars have a spectroscopic redshift.

In another recent paper \citep{georgakakis15}
the space density of distant quasars was also estimated
from the {XMM-XXL} survey data. The {XMM-XXL}
sky coverage area and the number of detected
quasars at $z>3$ are comparable to the sample by
\cite{kalfonzou14}.
Assembling the luminosity function sample \cite{georgakakis15} obtained spectroscopic
redshifts for most of the X-ray quasar candidates 
with apparent magnitudes $\rmag<22.5$ \citep{menzel16}. 
\mbox{Fig.~\ref{fig:densredshift} shows} (without errors) the estimates
of the space density of luminous quasars obtained
from the analytical X-ray luminosity function
model from \cite{georgakakis15} in the three
luminosity bins under consideration. The density estimates
in the luminosity range $44.5\leq\log\LX210<45.0$
from the data by \cite{georgakakis15} turn
out to be slightly lower than those obtained in this paper.
The estimates in the mutual luminosity range ($45.0\leq\log\LX210<45.3$) for both samples are 
in agreement.
Possible causes of the discrepancy are discussed in
the next section.

\section{DISCUSSION}

We were able to obtain a large sample (K16)
of sources at $z>3$ and luminosities \mbox{$\LX210>10^{45}$~erg/s}, i.e., above the break 
\mbox{($L_* \sim 5\times10^{44}$)}
in the X-ray luminosity function of quasars, and
to determine the slope $\gamma_2$ of the bright end of the
luminosity function (see Eq. (\ref{eq:2slope})). Since all of the
sources from the \K16\ subsample have luminosities
above $L_*$, they constrain the slope $\gamma_2$. In this case, it
should be kept in mind that the luminosities of many
of the \K16\ objects are higher than the presumed break
luminosity only by a factor $\gtrsim 2$, i.e., the region in
which the slope of the luminosity function changes
gradually from $\gamma_1$ to $\gamma_2$ could be touched.

To reliably determine all parameters of the luminosity
function, including $\gamma_2$, we supplemented
the \K16\ sample by another sample (\V14U) that includes
quasars with luminosities $\LX210 \lesssim L_*$. The
\V14U\ ($\LX210\lesssim 10^{45}$~erg/s) and \K16\ 
($\LX210>10^{45}$~erg/s) subsamples complement each other,
spanning virtually nonoverlapping luminosity ranges, but, at the same time, having a different completeness.

It follows from Table~\ref{tab:LFA1} that the parameters $A$, $\gamma_2$, $L_*$ depend on the incompleteness correction
more strongly than do the remaining ones. For the
listed parameters the bias in their values due to the
variations in the incompleteness correction turns out
to be larger than or comparable to their statistical
errors.

The beginning and the end of the bright slope of
the luminosity function is determined, respectively,
by the objects with $L>L_*$ from \V14U \ and the luminous
objects with \mbox{$L>2\times10^{45}$~erg/s} from the
\K16 \ subsample, whose the incompleteness correction
is close to unity. This reduces the uncertainty
in the slope $\gamma_2$ related to the \K16\ subsample
sources in the luminosity range \mbox{$10^{45}<L<2\times10^{45}$~erg/s}, for which the uncertainty in the
incompleteness correction is great. If $\gamma_2$ were determined
only with the \K16\ subsample, then its error and
the uncertainties in determining other parameters of
the luminosity function would be greater.

The bright end slope of the LDDE luminosity
function model and its statistical error are \mbox{$\gamma_2=\gamaod\pm\gamaoderr$} for incompleteness II. 
The uncertainty in the quasar detection completeness almost does not affect the slope value.

\subsection{Comparison of $\gamma_2$ with Previous Estimates}

Strictly speaking, our measured slope $\gamma_2$ of the X-ray
luminosity function of type~1 quasars cannot be
compared directly with the results of previous papers
\citep{vito14,ueda14,georgakakis15}, because the parameters of the luminosity
function models in them were obtained by taking
into account absorbed quasars. However, the statistics
of distant high-luminosity quasars ($\LX210>10^{45}$~erg/s) is usually based on large-area X-ray
surveys with shallow coverage in the X-ray and optical
bands. In such surveys the fraction of the unabsorbed
sources found is, as a rule, small (see, e.g.,
Kalfountzou et al. 2014). In this case, it should be
kept in mind that at a small number of X-ray counts it
is virtually impossible to distinguish a distant quasar
with $\NH\simeq 10^{23}$~cm$^{-2}$ from a quasar with a lower
absorption \citep{fotopoulou16}. Therefore, it can
be assumed that the published values of the slope
$\gamma_2$ are determined mainly by unabsorbed or weakly
absorbed sources with $\NH\lesssim 10^{23}$~cm$^{-2}$, generaly
by type 1 AGNs. This allows our estimate of the parameter
$\gamma_2$ to be approximately compared with the results
of other papers.

It is correct to compare the values of $\gamma_2$ only within
one empirical luminosity function model. Therefore,
for comparison with the results of \cite{vito14,georgakakis15}
we will choose our best LDDE model.

Samples of quasars characterized by a higher optical
identification completeness than \K16\ were used
in the papers chosen for our comparison. The sample
by \cite{vito14} consists of quasars at $z>3$
selected in the 0.5--2~кэВ band and was partially
used by us to supplement the \K16\ catalog by lower luminosity
objects. \cite{georgakakis15} studied
quasars at $\zspec>3$ selected in the 0.5-10 keV
band in the XMM-XXL survey region with an area
of 18~sq.~deg. A sample of 59 quasars at $\zspec>3$ was obtained through deep spectroscopic support
\citep{menzel16,georgakakis15} of this
region (deeper than on average for SDSS by 2 magnitudes).
The XMM-XXL survey was supplemented
by data from deep Chandra X-ray surveys (CDFS,
CDFN, AEGIS, ECDFS, and C-COSMOS) spanning
the luminosity range \mbox{$10^{43}$--$10^{45}$}~erg/s and
yielded significant estimates of the density of quasars
at luminosities $>10^{45}$~erg/s. Therefore, the results
of \cite{georgakakis15} turn out to be most
interesting for our comparison.

\begin{figure}[!ht]
\centering
\includegraphics[width=0.75\linewidth]{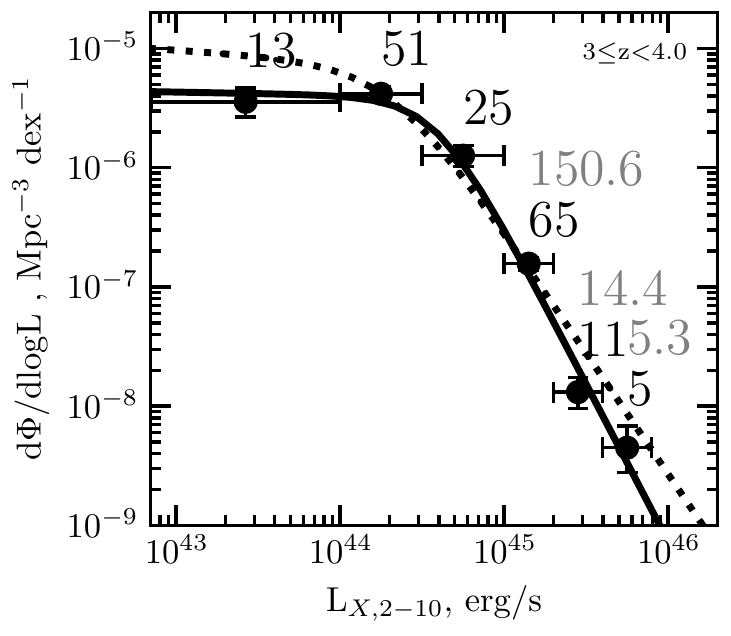}
\includegraphics[width=0.75\linewidth]{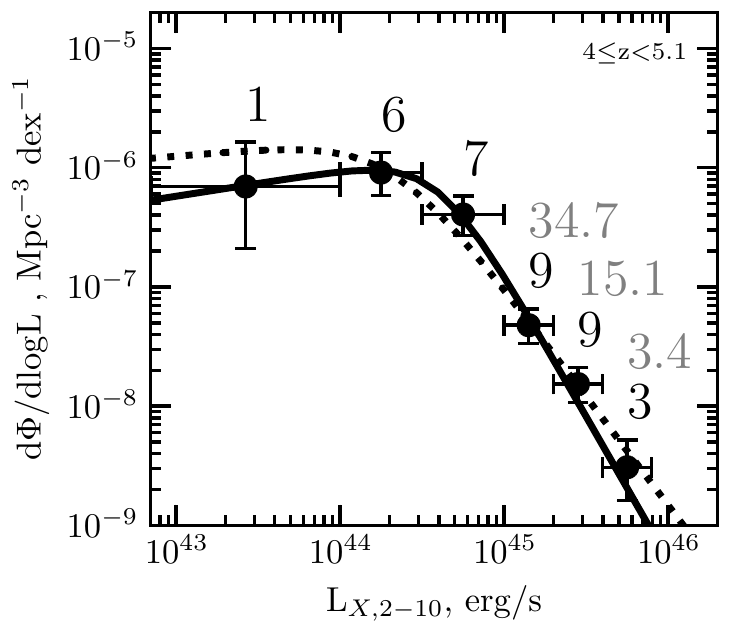}

\caption{The differential luminosity function in two broad redshift intervals. 
The points with error bars indicate the binned luminosity 
function obtained by $1/\Vmax$ method using the incompletenes correction~II.
The black font indicates the number of sample objects in the $\Delta\log L$--$\Delta z$
bins under consideration, while the gray font indicates the expected total number of objects
corrected for the \K16 \ subsamble incompleteness~II. The black solid line represents the LDDE model computed for the incompleteness correction II. The dashed line indicates the LDDE model from \cite{georgakakis15}. 
} 
\label{fig:comparasion}
\end{figure}

The bright end slope value of the luminosity
function $\gamma_2 =\gamaod \pm \gamaoderr$ obtained in our paper for
the LDDE model and the incompleteness correction~II intersects the $1\sigma$ confidence interval of the estimates
by \cite{vito14} for the LDDE model, $\gamma_2 =3.71_{-0.84}^{+1.12}$.
The $\gamma_2$ measurement accuracy improved
significantly compared to deep small-area surveys
\citep{vito14}. However, there is disagreement
with the results from \cite{georgakakis15},
where a considerably smaller slope was derived for
the LDDE model, $\gamma_2=2.15\pm0.24$ (see a comparison
of the luminosity functions derived in our paper and
\cite{georgakakis15} in Fig.~\ref{fig:comparasion}). The luminosity
break $L_*=44.31 \pm 0.13$\footnote{The cosmological parameters $\Omega_m$ and $\Omega_\lambda$ in our paper and
\cite{georgakakis15} slightly differ.} from \citep{georgakakis15}
is also lower than our estimates (see Table~\ref{tab:LFA1}).

The estimate of the density of quasars in the range
$44.7<\log\LX210<45.3$ from the \K16\ subsample
turns out to be differ than follows from the \mbox{model
by \cite{georgakakis15}, see Fig.~\ref{fig:densredshift}}. However,
it follows from Fig.~\ref{fig:comparasion} that the difference between the
models is not that significant and they agree between
themselves, within the statistical error limits. 
A discrepancy of the density at the highest luminosity range where \K16\ has objects is interesting for further research.

The difference in the estimates of $\gamma_2$ and $L_*$ under
discussion can be caused by the following factors.
First, at luminosities $\LX210 \lesssim 5\times10^{44}$~erg/s
sources from deep surveys appear in the sample by
\cite{georgakakis15}, and their contribution
changes significantly the density distribution with
respect to unabsorbed quasars. That is why the
points belonging the \V14U\ subsample in Fig.~\ref{fig:comparasion} lie well below the model by \cite{georgakakis15}.

Second, the area of the deep surveys used in \cite{georgakakis15} is half the area of the deep surveys
from \cite{vito14} used in our paper. Consequently,
in the sample by \cite{vito14} there
are more objects with luminosities near the break
luminosity $L_*\simeq5 \times 10^{44}$~erg/s \citep{vito14}
than in the deep surveys of the sample by \cite{georgakakis15}, 
and, therefore, the sample by \cite{vito14} allows $L_*$ to be determined more accurately.

Third, in contrast to our paper and \cite{vito14},
in which a certain value of photometric redshifts $\zphot$ were
assigned to the objects, a probabilistic approach was
used in \cite{georgakakis15}: the probability density distribution of
possible $\zphot$ was сonsidered for each object from the
deep surveys. \cite{georgakakis15} showed that
using fixed $\zphot$ in analyzing the data of deep surveys,
such as COSMOS, could lead to an overestimate (by
a factor of 1.8--3) of the density of quasars at luminosities
$\lesssim 5\times 10^{44}$~erg/s. 
Therefore, some of the
photometric candidates from the \V14U\ subsample
may turn out to be quasars at lower $z$ and the slope
$\gamma_2$ will then be shallower. 
The slope~$\gamma_2$ can be overestimated
if type~2 quasars, without broad lines in the
optical spectrum (see Fig.~\ref{fig:monochromatic}), are present among the
photometric quasar candidates with $\NH \leqslant 10^{23}$~cm$^{-2}$
from the \V14U\ subsample. In the range $44.5<\log\LX210<45.0$, which defines the beginning of
the slope~$\gamma_2$, the fraction of photometric candidates
in the \V14U\ subsample is about 20\%. Therefore, if
there are absorbed quasars or quasars $z<3$ among the $\zphot$ candidates, 
then this will not affect strongly
the estimate of the slope~$\gamma_2$.

Fourth, the spectroscopic sample by \cite{menzel16} used in \cite{georgakakis15} may
be subjected to optical identification incompleteness
at 0.5--2 keV X-ray fluxes $\lesssim 5\times10^{-15}$~erg/s/cm$^2$
corresponding to luminosities$\sim 5\times10^{44}$~erg/s for
quasars at $z>3$. In that case, the measurements will
show a shallower slope $\gamma_2$ than the actual one.

All of the listed factors can lead to a mismatch between
the values of $L_*$ and $\gamma_2$ that were obtained in this paper and \cite{georgakakis15}.

\begin{figure}[!ht]
\centering
\includegraphics[width=0.8\linewidth]{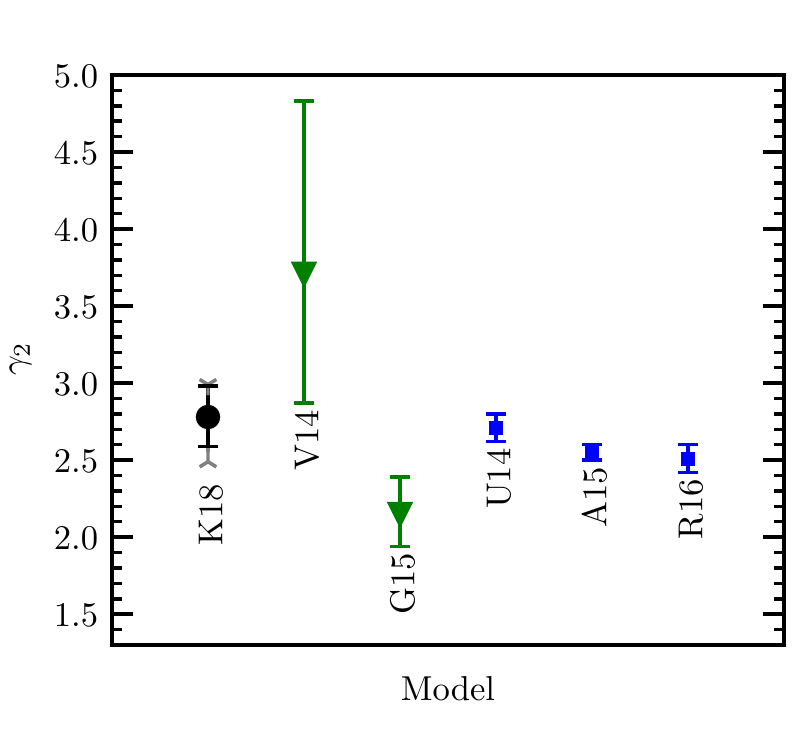}
\caption{Comparison of the bright end slope ($\gamma_2$) obtained in this paper (K18) with the values reported before, for the LDDE luminosity function model. The black circle with error bars indicates the slope value and its statistical uncertainty $\gamma_2=\gamaod\pm \gamaoderr$ for the LDDE model, for incompleteness correction II. The gray error bars show the range of $\gamma_2$ values for incompleteness corrections I and III, taking into account statistical errors. The green triangles denote the LDDE $\gamma_2$ values and the corresponding statistical errors obtained for quasar samples at $z>3$: V14 \citep{vito14}; G15 \citep{georgakakis15}. The blue squares show the LDDE $\gamma_2$ estimates obtained for large quasar samples spanning a wide range of luminosities and redshifts, in which high-redshift objects at $z>3$ constitute just a few percent of the total sample: U14 \citep{ueda14}; A15 \citep{aird15} --- a soft-band sample; R16 \citep{ranalli16}.}
\label{fig:gammacomp}
\end{figure}

In Fig.~\ref{fig:gammacomp} the values of the slope $\gamma_2$ derived in
our paper are compared with the results of \cite{vito14,georgakakis15}. It
can be clearly seen that using the \K16\ objects that
we selected based on the data of a large-area X-ray
survey, we were able to constrain the slope of
the bright end of the X-ray luminosity function for
distant quasars much better than can be done based
only on the data of small-area deep surveys \citep{vito14}. 
The same figure shows the values of
$\gamma_2$ from \cite{ueda14,aird15,ranalli16}, where the luminosity function
models were constructed based on samples of quasars
spanning a wide range of luminosities and redshifts.
In these papers the quasars at $z > 3$ account for
only a few percent of the total size of the samples,
which consist of absorbed and unabsorbed quasars;
in addition, more complex luminosity function models
dependent on a larger number of evolution parameters
were used. Therefore, the values of $\gamma_2$ obtained in
these papers characterize the distribution of more
nearby quasars.

\section{CONCLUSIONS}

In this paper we obtained estimates of the \mbox{X-ray} luminosity
function for type~1 quasars for a sample of
101 sources with luminosities $L_{X,2-10} \geqslant 10^{45}$~erg/s
from our catalog  \citep{khorunzhev16}. The LDDE,
LADE, ILDE, and PDE luminosity function models
describe equally well the density distribution of unabsorbed quasars.
The constraints on the bright end slope of the X-ray luminosity
luminosity function ($\gamma_2 =\gamaod\pm \gamaoderr$ for the
LDDE model) were improved.

The values of $\gamma_2$ and other model parameters depend
on the choice of a quasar incompleteness correction
for the \K16\ catalog. As the correction increases,
the slope $\gamma_2$ becomes steeper and the break
luminosity grows.

The necessity of taking into account this correction
stems from the fact that only for sources with
$\zmag<20.5$ we can make photometric redshift estimates
using the entire set of SDSS filters, thus improving
the reliability and accuracy of $\zphot$. In this case, some
of the X-ray luminous quasars at $z>3$ turn out to be
fainter than the chosen optical threshold and will be
missed in the selection.

Most of the \K16 \ sources selected by $\zphot$ are
spectroscopically confirmed SDSS quasars. The
sample of distant X-ray quasars at luminosities
$L_{X,2-10}>10^{45}$~erg/s can be expanded by 20\%
by the method of searching for new candidates for
distant quasars described in \cite{khorunzhev16}.
These candidates are confirmed by the spectroscopic
observations performed at the following telescopes:
\AZT33IK\ \citep{kamus02} with the ADAM
low-resolution spectrograph \citep{afanasev16,burenin16} and \BTA\ with the SCORPIO-I
\citep{afanasev05} and SCORPIO-II
\citep{afanasev11,afanasev12} spectrographs (see~\cite{khorunzhev17,khorunzhev17b,khorunzhev18b}).

The produced X-ray sample of luminous quasars
at $z>3$ is one of the most extensive in sky coverage
area and number of luminous sources. It can be
used as a reference one to estimate the completeness
and purity of the methods for the selection of distant
quasars and to test the algorithms for optical identifications
of X-ray sources from the planned SRG all-sky
survey \citep{pavlinsky11,merloni14}.

An X-ray quasar at $z=3$ with a \mbox{0.5--2~keV} flux
$\simeq10^{-14}$~erg/s/cm$^2$ has a 2--10 keV luminosity
$\LX210\simeq 10^{45}$~erg/s. This means that $\zphot$ in
SDSS fields can be obtained for $\gtrsim 50$\% of the X-ray
quasars at $z\sim3$ found in the SRG/eROSITA survey
\citep{merloni14} with fluxes $\gtrsim 10^{-14}$~erg/s/cm$^2$,
which corresponds to the average sensitivity of a
four-year survey over the sky. It will be possible to
refine the break luminosity \mbox{($L_*\simeq 4\times10^{44}$~erg/s)}
using the data of deep SRG survey fields near the
poles of the ecliptic, where a sensitivity $\FX0.52\simeq 2\times10^{-15}$~erg/s/cm$^2$ will be achieved.

We are planning to expand the existing sample of
distant X-ray quasars through new X-ray (\XMM)
and optical (\SDSS, Pan-STARRS) data,
to improve the selection methods (see, e.g., \citealt{mescherakov15}, and to continue the program of
their spectroscopic identification with the AZT-33IK
and BTA telescopes.

\section{ACKNOWLEDGMENTS}
This study was supported by RSF \mbox{(project {No. 14-22-00271})}. 
The observations at the 6-m BTA telescopes
were financially supported by the Ministry
of Education and Science of the Russian Federation
(contract no. 14.619.21.0004, project identifier
RFMEFI61914X0004).
The AZT-33IK observations were done by using the equipment of Center for Common Use
"Angara"\ \url{http://ckp-rf.ru/ckp/3056/}. 
The working capacity of AZT-33IK equipment was supported by funding of Basic Research program II.16.
We would like to thank V.~Astakhov for translation of the paper in English.

\vfill
\eject

\onecolumn
\newcommand\oldtabcolsep{\tabcolsep}
\setlength{\tabcolsep}{2pt}

{\pagestyle{empty}

\begin{longtable}{rrccrrlcccc}

\caption{The sample of sources from the \K16\ catalog with $\LX210>10^{45}$ erg/s}\label{tab:catlum}\\
\hline
\hline
\multicolumn{1}{c}{N} &
\multicolumn{1}{c}{No} &
\multicolumn{1}{c}{Name} &
\multicolumn{1}{c}{OBJID} &
\multicolumn{1}{c}{RA} &
\multicolumn{1}{c}{DEC} &
\multicolumn{1}{c}{$z$} &
\multicolumn{1}{c}{$z_{\rm ref}$} &
\multicolumn{1}{c}{$\FX0.52$} &
\multicolumn{1}{c}{$\LX210$}
\\
\hline
\endhead 
\hline 
\hline
\caption*{
\scriptsize
\raggedright
\textbf{Notes:} N is the source number, No is the ordinal source number in the catalog (\K16) \cite{khorunzhev16}, 
the asterisk <<*>> marks sources have being taken from the additional table of quasars with $\zspec>3$
that did not enter into the catalog of candidates in the paper \cite{khorunzhev16}, 
Name --- is the name in the \3XMMDR4 (3XMMJ...) catalog, OBJID --- is the identifier in the photometric \SDSS--{DR12} catalog, 
RA and DEC are the right ascension and declination (\SDSS--{DR12}) in degrees, $z$ is the redshift of the source, 
$z_{\rm ref}$ is a reference to the redshift: 
the empty field is the photometric redshift \citep{khorunzhev16}, 
1 is the \SDSS--{DR12} spectroscopy \citep{alam15}, 2 is the \AZT33IK и \BTA \ spectroscopy \citep{khorunzhev17,khorunzhev17b,khorunzhev18b}; the remaining redshifts were taken from the catalog \citealt{flesch15} (the reference numbers correspond to \cite{flesch15}: 611 --- \cite{flesch15}, 643 --- \citep{gandhi02},  646 --- \cite{garilli14},  1297 --- \cite{monier02}, 1347 --- \cite{newman13}, 1406 --- \cite{papovich06},  1411 --- \cite{paris17}, 1758 --- \cite{stalin10}), $\FX0.52$ is the 0.5--2~keV X-ray \mbox{flux~$\times10^{-14}$~erg/s/cm$^2$}, 
$\LX210$ is the decimal logarithm of the source’s 2--10~keV luminosity (erg/s) in its rest frame.
}
\endfoot
1&5&J000443.6$-$084036&1237680240914071885&1.1820&-8.6761&3.85&&0.94&45.21\\ 
2&8&J000618.1$-$084410&1237672793424200167&1.5758&-8.7359&3.323&1&0.87&45.03\\ 
3&27&J002706.9+261559&1237680275262538220&6.7800&26.2667&3.29&&0.94&45.06\\ 
4&30&J003000.5+044040&1237678661427266242&7.5034&4.6784&3.091&1411&1.00&45.02\\ 
5&35&J004054.6$-$091527&1237652948530037577&10.2277&-9.2575&5.002&1&1.11&45.53\\ 
6&42&J004505.3$-$014048&1237678881562427510&11.2721&-1.6800&3.282&1&1.00&45.08\\ 
7&45&J004800.9+315354&1237680310696804736&12.0039&31.8986&3.18&&1.40&45.19\\ 
8&50&J005952.7+314403&1237680310697919062&14.9693&31.7343&3.30&&1.84&45.35\\ 
9&87&J020229.4$-$042703&1237679323396309357&30.6225&-4.4509&3.23&&2.11&45.39\\ 
10&89&J020316.4$-$074831&1237679338956325563&30.8182&-7.8090&3.296&1&1.24&45.18\\ 
11&107&J021126.4$-$054022&1237679321786614354&32.8598&-5.6731&3.399&1&0.99&45.11\\ 
12&115&J021401.9$-$003941&1237663783138296681&33.5082&-0.6617&4.17&&0.63&45.11\\ 
13&133&J022037.4$-$061037&1237679340568903780&35.1561&-6.1769&3.03&&1.01&45.01\\ 
14&141&J022112.5$-$034251&1237679323935212347&35.3026&-3.7145&5.011&1&0.62&45.28\\ 
15&144&J022307.9$-$030840&1237679255745790580&35.7832&-3.1445&3.675&1&0.77&45.07\\ 
16&153&J022320.7$-$031823&1237678887988429287&35.8363&-3.3068&3.865&1&2.19&45.58\\ 
17&163&J022826.5$-$085501&1237652900227252760&37.1099&-8.9175&3.24&&1.17&45.13\\ 
18&167&J022906.0$-$051428&1237679253062091149&37.2752&-5.2414&3.173&1&1.93&45.33\\ 
19&180&J023441.1$-$040711&1237679323399782556&38.6713&-4.1197&3.334&1&0.95&45.07\\ 
20&192&J030449.8$-$000814&1237666300553789504&46.2077&-0.1371&3.287&1&4.81&45.76\\ 
21&245&J084617.8+190342&1237667211581522773&131.5738&19.0620&3.47&2&0.99&45.13\\ 
22&257&J085822.2+564533&1237660936091796090&134.5925&56.7590&3.021&1&1.34&45.12\\ 
23&282&J091959.5+370550&1237660634915406290&139.9984&37.0974&3.379&1&0.80&45.01\\ 
24&286&J092143.5+063644&1237658425155977396&140.4313&6.6121&3.718&1&1.00&45.20\\ 
25&287&J092325.3+453223&1237657401346424982&140.8552&45.5395&3.452&1&1.49&45.30\\ 
26&292&J093404.6+472434&1237657590848618536&143.5195&47.4095&3.086&1&1.74&45.26\\ 
27&293&J093451.6+353744&1237661384382480820&143.7148&35.6290&3.363&1&0.96&45.08\\ 
28&296&J093709.6+495147&1237657770707976723&144.2908&49.8642&3.641&1411&3.00&45.66\\ 
29&318&J095937.0+131212&1237664106852384915&149.9046&13.2043&4.064&1411&1.88&45.56\\ 
30&338&J101515.2+085456&1237660584444953274&153.8140&8.9159&3.235&1&1.46&45.23\\ 
31&347&J102107.5+220922&1237667538009588107&155.2816&22.1560&4.262&1&1.74&45.57\\ 
32&370&J103428.8+393343&1237661383314178468&158.6203&39.5621&4.334&1411&1.38&45.49\\ 
33&382&J104612.9+584719&1237655109446467756&161.5541&58.7886&3.054&1&1.53&45.19\\ 
34&385&J104909.8+373758&1237664668437774491&162.2909&37.6331&3.005&1&6.95&45.83\\ 
35&396&J105049.2+354517&1237664819280347214&162.7057&35.7557&3.326&1411&0.88&45.04\\ 
36&398&J105123.0+354535&1237664819280412861&162.8460&35.7595&4.921&1&1.73&45.71\\ 
37&411&J110458.2+250421&1237667551956369534&166.2428&25.0728&3.522&1&1.80&45.40\\ 
38&430&J111900.0+152707&1237661070867431568&169.7508&15.4520&3.138&1&1.12&45.08\\ 
39&431&J112020.9+432545&1237661850390954212&170.0874&43.4292&3.555&1&0.88&45.10\\ 
40&447&J114323.7+193447&1237667915416600770&175.8488&19.5800&3.348&1&1.19&45.17\\ 
41&449&J114447.7+370434&1237664818748260677&176.1986&37.0763&4.010&1&1.77&45.52\\ 
42&453&J114816.0+525900&1237657857682899337&177.0670&52.9831&3.173&1&6.74&45.87\\ 
43&459&J115839.8+262510&1237667429035869276&179.6659&26.4197&3.428&1&0.94&45.09\\ 
44&460&J115933.3+553632&1237657591395844307&179.8888&55.6091&3.981&1&0.58&45.03\\ 
45&463&J120125.5+064621&1237671140947592014&180.3563&6.7729&3.323&1&1.24&45.18\\ 
46&476&J120949.7+453400&1237661873476534381&182.4573&45.5668&3.609&1&2.72&45.61\\ 
47&510&J122602.0+132114&1237661813886091391&186.5088&13.3540&3.530&1&1.95&45.44\\ 
48&523&J123136.8+131544&1237661950792696231&187.9030&13.2617&3.48&2&0.99&45.13\\ 
49&524&J123005.9+142957&1237664289929494661&187.5244&14.4989&3.275&1&1.56&45.27\\ 
50&525&J123011.9+102237&1237662238004412598&187.5500&10.3771&3.569&1&1.00&45.16\\ 
51&529&J123157.3+000933&1237648704579108962&187.9891&0.1590&3.226&1&0.95&45.04\\ 
52&538&J123503.1$-$000331&1237648721234559206&188.7627&-0.0588&4.701&1&2.27&45.78\\ 
53&553&J124210.7+024049&1237671765324595744&190.5448&2.6804&3.175&1&3.12&45.54\\ 
54&569&J125736.2+242040&1237667911133888887&194.4003&24.3444&3.681&1&0.92&45.15\\ 
55&579&J130616.9+264335&1237667322724680051&196.5703&26.7264&3.208&1&1.21&45.14\\ 
56&580&J130811.9+292512&1237665428627456150&197.0497&29.4202&3.035&1&1.11&45.04\\ 
57&592&J131236.2+231629&1237667910061654088&198.1511&23.2751&3.684&1&0.65&45.00\\ 
58&618&J133200.0+503613&1237662301357736036&202.9998&50.6037&3.84&2&0.78&45.12\\ 
59&619&J133223.2+503430&1237662301357736105&203.0969&50.5754&3.832&1&0.94&45.20\\ 
60&627&J134135.6$-$001321&1237648704049840848&205.3980&-0.2230&3.919&1&0.67&45.07\\ 
61&653&J140146.5+024433&1237651754560520506&210.4439&2.7430&4.424&1&1.06&45.39\\ 
62&675&J142926.4+011951&1237651752952923130&217.3601&1.3316&4.840&1297&0.51&45.17\\ 
63&693&J145753.0$-$011358&1237648702984422397&224.4710&-1.2330&3.503&1&1.33&45.26\\ 
64&704&J151147.1+071406&1237662237485039775&227.9465&7.2350&3.481&1&2.58&45.55\\ 
65&710&J151534.3$-$000000&1237648721252122996&228.8933&-0.0002&3.04&2&1.38&45.14\\ 
66&731&J154905.8+352020&1237662503219364016&237.2744&35.3390&3.038&1&2.60&45.42\\ 
67&745&J160528.3+272852&1237662307273999256&241.3675&27.4818&4.023&1&0.77&45.16\\ 
68&755&J162114.9$-$021130&1237668651464918353&245.3125&-2.1918&4.34&2&0.67&45.17\\ 
69&762&J163207.9+571108&1237668505439503219&248.0339&57.1863&3.40&2&1.04&45.13\\ 
70&766&J163459.2+332510&1237661386008298072&248.7476&33.4194&3.237&1&1.36&45.20\\ 
71&782&J171337.2+585306&1237651225708921950&258.4049&58.8853&4.37&2&0.76&45.24\\ 
72&796&J203958.0$-$004337&1237656567574104067&309.9923&-0.7273&4.63&&0.44&45.06\\ 
73&816&J212959.5+051005&1237669762254439608&322.4981&5.1683&3.02&&1.01&45.00\\ 
74&826&J215139.1+021628&1237678597539561948&327.9136&2.2740&3.256&1&0.97&45.06\\ 
75&837&J221753.2$-$003257&1237663542611083691&334.4730&-0.5486&3.106&1411&1.69&45.25\\ 
76&840&J222008.9$-$002343&1237663478722658939&335.0375&-0.3955&3.344&1&0.89&45.05\\ 
77&856&J230252.1+085522&1237679034548486973&345.7172&8.9225&3.750&1&0.64&45.02\\ 
78&859&J231619.4+254552&1237666184031633742&349.0811&25.7647&3.207&1&1.01&45.06\\ 
79&863&J231839.7+002032&1237666408437907970&349.6655&0.3421&3.23&&1.11&45.11\\ 
80&866&J232137.4+283025&1237680331636474144&350.4056&28.5072&3.062&1&1.67&45.23\\ 
81&871&J232346.0+165228&1237678601301459610&350.9415&16.8744&3.602&1&1.22&45.25\\ 
82&872&J232419.4+165620&1237678601301524724&351.0810&16.9389&3.323&1&1.50&45.27\\ 
83&890&J234214.1+303606&1237666183498039666&355.5590&30.6017&3.37&2&0.80&45.01\\ 
84&897&J235054.6+200939&1237680246813491428&357.7276&20.1607&3.162&1&1.00&45.04\\ 
85&898&J235201.3+200901&1237680246813556916&358.0054&20.1507&3.079&1&1.16&45.08\\ 
86&901&J235435.5$-$101513&1237652900210671714&358.6483&-10.2537&3.120&1&1.16&45.09\\ 
87&*2&J002654.9+171944&1237678601308078496&6.7290&17.3290&3.095&1&1.12&45.07\\ 
88&*6&J020231.1$-$042246&1237679323396309664&30.6298&-4.3797&4.270&1&1.28&45.44\\ 
89&*9&J021338.6$-$051615&1237679253060387565&33.4110&-5.2711&4.544&1&0.93&45.36\\ 
90&*13&J022251.7$-$050713&1237679322324795732&35.7157&-5.1202&3.860&1758&0.88&45.18\\ 
91&*17&J023226.0$-$053729&1237679341107085527&38.1089&-5.6249&4.564&1&0.53&45.12\\ 
92&*24&J093521.2+612339&1237651272966275457&143.8391&61.3942&4.042&1&0.69&45.12\\ 
93&*25&J094013.9+344628&1237661382772130308&145.0579&34.7747&3.355&1&2.30&45.46\\ 
94&*30&J100655.8+050325&1237658297920454886&151.7325&5.0569&3.086&1&1.45&45.18\\ 
95&*33&J104808.3+583718&1237658304353272305&162.0354&58.6210&3.285&1&1.14&45.13\\ 
96&*43&J124405.1+125757&1237661817633374639&191.0211&12.9658&3.100&611&1.10&45.06\\ 
97&*51&J140149.8+024835&1237651754560520571&210.4579&2.8102&3.830&643&0.98&45.22\\ 
98&*54&J150603.5+012757&1237651753493791548&226.5146&1.4662&3.852&1&0.79&45.13\\ 
99&*56&J164829.7+350159&1237659326568858151&252.1238&35.0330&4.075&1347&3.53&45.84\\ 
100&*57&J171456.2+593700&1237651226245530116&258.7344&59.6169&4.028&1406&1.20&45.36\\ 
101&*61&J220845.5+020252&1237678597004591287&332.1895&2.0479&3.405&646&1.00&45.12\\ 
\hline
\hline

\end{longtable}

\setlength{\tabcolsep}{\oldtabcolsep}
} 

\onecolumn
\begin{landscape}
\centering

\centering
\scriptsize
\setlength{\tabcolsep}{2pt}
\begin{longtable}{cccccccccc}
\caption{Parameters of the X-ray luminosity function models for the \V14U\ and \K16 \ subsamples.} \label{tab:LFA1}\\ 
\hline
\hline
\multicolumn{1}{c}{Модель} &
\multicolumn{1}{c}{$\log A$} &
\multicolumn{1}{c}{$\log L_*$} &
\multicolumn{1}{c}{$\gamma_1$} &
\multicolumn{1}{c}{$\gamma_2$} &
\multicolumn{1}{c}{$\plum$} &
\multicolumn{1}{c}{$\pden$} &
\multicolumn{1}{c}{$\beta$} &
\multicolumn{1}{c}{\tiny $\Delta AIC$} &
\multicolumn{1}{c}{\tiny $\Delta BIC$}
\\
\hline
\endhead 
\hline 
\endfoot
PDE&-5.15$^{+0.11(-0.11)}_{-0.11(0.08)}$ &44.58$^{+0.07(0.15)}_{-0.08(-0.09)}$ &0.05$^{+0.15(0.09)}_{-0.17(-0.07)}$ &2.59$^{+0.18(-0.03)}_{
-0.17(0.00)}$ & --- &-4.83$^{+0.68(0.20)}_{-0.70(-0.27)}$ & --- &2.3 &0.0\\ 
ILDE&-5.04$^{+0.12(-0.11)}_{-0.13(0.06)}$ &44.51$^{+0.08(0.14)}_{-0.09(-0.08)}$ &0.03$^{+0.16(0.12)}_{-0.18(-0.08)}$ &2.62$^{+0.18(0.03)}_{
-0.17(-0.01)}$ &1.35$^{+0.70(0.58)}_{-0.70(-0.31)}$ &-7.04$^{+1.34(-0.35)}_{-1.36(0.11)}$ & --- &0.6 &1.7\\ 
LADE&-5.06$^{+0.12(-0.11)}_{-0.12(0.06)}$ &44.51$^{+0.08(0.14)}_{-0.09(-0.08)}$ &0.03$^{+0.16(0.12)}_{-0.18(-0.08)}$ &2.61$^{+0.18(0.03)}_{
-0.17(-0.01)}$ &1.38$^{+0.71(0.56)}_{-0.70(-0.30)}$ &-0.65$^{+0.12(-0.03)}_{-0.12(0.01)}$ & --- &0.3 &1.3\\ 
LDDE&-5.13$^{+0.12(-0.13)}_{-0.12(0.08)}$ &44.59$^{+0.07(0.14)}_{-0.07(-0.09)}$ &0.16$^{+0.16(0.11)}_{-0.18(-0.09)}$ &2.78$^{+0.20(0.00)}_{
-0.19(-0.04)}$ & --- &-6.95$^{+1.27(-0.24)}_{-1.33(0.08)}$ &2.64$^{+1.31(0.57)}_{-1.28(-0.45)}$ &0.0 &1.0\\ 
PLE&-5.40$^{+0.12(-0.10)}_{-0.12(0.07)}$ &44.64$^{+0.08(0.14)}_{-0.09(-0.09)}$ &0.10$^{+0.15(0.08)}_{-0.17(-0.06)}$ &2.43$^{+0.17(-0.08)}_{
-0.16(0.02)}$ &-2.10$^{+0.39(0.01)}_{-0.41(-0.12)}$ & --- & --- &24.6 &22.4\\

\hline
\hline
\caption*{
\scriptsize
\raggedright
\textbf{Notes:} LDDE, ILDE, LADE, PDE, PLE are the models under consideration, 
$\log A$ is the decimal logarithm of the normalization factor (Мpc$^{-3}$),
$\log L_*$ is the decimal logarithm of the break luminosity (erg/s),
$\gamma_1$ and $\gamma_2$ are the exponents for the faint and bright slopes of the luminosity function, 
$\plum$ is the luminosity evolution parameter,
$\pden$ is the density evolution parameter, 
$\beta$ is an additional parameter that accounts for the luminosity 
dependency of the LDDE model, $\Delta{AIC}$ and 
$\Delta{BIC}$ are the differences of the AIC and BIC
information criteria for a
given model and the model with the lowest AIC (LDDE) и BIC (PDE) values . 
The parameters and their statistical errors
(1$\sigma$) are given for
the incompleteness correction~II. The shift of the parameters for the incompleteness corrections~I~and~III 
relative to the incompleteness
correction~II is given in parentheses at the bottom and the top, respectively.
}
\endlastfoot
\end{longtable}

\captionsetup{width=12cm}
\begin{longtable}{cccccc}
\caption{The number of sources in $\Delta\log
  L$--$\Delta z$ bins} \label{tab:bindistr} \\  
\hline
\hline
\multicolumn{1}{c}{$\Delta\log L$/$\Delta z$} &
\multicolumn{1}{c}{3.00--3.19} &
\multicolumn{1}{c}{3.19--3.47} &
\multicolumn{1}{c}{3.47--3.90} &
\multicolumn{1}{c}{3.90--4.30} &
\multicolumn{1}{c}{4.30--5.10} \\
\hline
\endhead
\hline
\caption*{
\scriptsize
\raggedright
\textbf{Notes:} The rows in the table show the binning by logarithm of the X-ray luminosity~$\LX210$. The columns show the binning by redshift $z$. In the cells of the table the number of objects in a given bin from the \V14U \ and \K16\ samples is specified on the left and the right,
respectively.
} 
\endfoot
42.9-44.0&5/-&4/-&4/-&1/-&-/-\\ 
44.0-44.5&18/-\ &17/-\ &15/-\ &5/-&2/-\\ 
44.5-45.0&6/-&7/-&11/-\ &5/-&3/-\\ 
45.0-45.3&\ -/18&\ 1/27&\ -/17&-/5&-/6\\ 
45.3-45.9&-/5&-/5&-/6&-/6&-/6\\ 
\hline
\end{longtable}

\end{landscape}

\twocolumn

\end{document}